\documentclass[preprint,12pt]{elsarticle}

\usepackage{amssymb}
\usepackage{amsmath,amsfonts}
\usepackage{xcolor}
\usepackage{booktabs}
\usepackage{siunitx}
\usepackage{tikz}
\usepackage{graphicx}
\usepackage{float}
\usepackage{algorithm}
\usepackage[percent]{overpic} 
\usepackage{amsmath}          
\usepackage{algpseudocode}
\usepackage{mathtools}
\usepackage{bm}
\usepackage{stmaryrd}
\usetikzlibrary{patterns, decorations.pathmorphing}

\usetikzlibrary{arrows.meta, positioning}
\journal{}

\begin{document}

\begin{frontmatter}



\title{Efficient Estimation of A-basis and B-Basis Value  under Epistemic Uncertainty using Importance Sampling and Control Variates}

\author[AIRBUS,ONERADTIS]{Elton Donfack-Siewe} 
\author[ONERADTIS]{Jérôme Morio}
\author[ONERADTIS]{Sylvain Dubreuil}
\author[AIRBUS]{Jean-Philippe Navarro}
\author[ONERADMAS]{Christian Fagiano}

\affiliation[AIRBUS]{organization={AIRBUS OPERATIONS SAS},
      addressline={316 route de Bayonne}, 
      city={Toulouse},
      postcode={31060}, 
      country={France}}
\affiliation[ONERADTIS]{organization={ONERA/DTIS},
      addressline={Université de Toulouse}, 
      city={Toulouse},
      postcode={31055}, 
      country={France}}
\affiliation[ONERADMAS]{organization={ONERA/DMAS},
      addressline={Université Paris Saclay}, 
      city={Châtillon},
      postcode={92320}, 
      country={France}}

\begin{abstract}
In aerospace certification and other safety-critical domains, conservative quantile estimation such as A- and B-basis values is essential to guarantee reliability. While these metrics are traditionally derived from experimental campaigns, this work focuses on their estimation using a validated deterministic numerical model. The problem is formulated under  mixed aleatory-epistemic uncertainty, accounting for limited material data, finite sampling effects, and surrogate modeling errors. We propose a methodology for estimating conservative design quantiles with statistical guarantees under mixed uncertainties. The proposed method leverages importance sampling and control variates to achieve accurate and efficient estimates within a fixed computational budget. One key point is the surrogate model's role solely as a variance reduction device, which guarantees unbiased and consistent quantile estimation. By explicitly integrating all sources of uncertainty, the proposed framework provides a numerical alternative to estimate A-basis and B-Basis. Furthermore, Sobol-based sensitivity indices are obtained at no additional cost, offering insight into the dominant epistemic sources. Numerical experiments on structural models demonstrate the method’s reliability and computational efficiency. In particular, the application to large-scale industrial simulations confirms its suitability for aerospace certification workflows and highlights its relevance for real-world engineering environments.
\end{abstract}
\begin{graphicalabstract}

\end{graphicalabstract}
\begin{highlights}
    \item Quantile estimation in the presence of three sources of epistemic uncertainty
    \item Propagation of epistemic and aleatory uncertainty with importance sampling
    \item Control variates for variance reduction and uncertainty-aware conservatism
    \item Sensitivity analysis on epistemic uncertainties
\end{highlights}


\end{frontmatter}




\section{Introduction}
Ensuring the reliability of complex engineering systems under uncertainty is a cornerstone of robust and certifiable design, particularly in safety-critical fields such as aerospace. In such contexts, engineers must guarantee that performance requirements are satisfied with high confidence, despite limited experimental data and limited numerical data (due to the high cost of numerical solvers). This challenge naturally calls for a rigorous framework of uncertainty quantification.

In most real-world applications, the system response is obtained from computationally expensive simulations, while input parameters are uncertain due to manufacturing variability, modeling approximations, or incomplete characterization. The task is to estimate a quantity of interest—such as a mean, a failure probability, or more specifically, a conservative quantile (e.g., A- or B-basis value)—by propagating these uncertainties through the numerical model.

Conventional approaches often rely on Monte Carlo sampling based on assumed probabilistic input models \cite{Rubinstein2016}. Variance-reduction techniques such as Importance Sampling (IS) \cite{TokdarKass2009}, Control Variates (CV) \cite{Hickernell2003}, stratified sampling \cite{Samawi2019}, and Conditional Monte Carlo \cite{Nakayama2014} have been proposed to improve estimator efficiency. However, when dealing with high-fidelity simulations, the computational cost required to obtain statistically reliable bounds remains prohibitive.

Uncertainty quantification distinguishes two complementary forms of uncertainty: aleatory uncertainty, reflecting inherent variability in the system, and epistemic uncertainty, arising from limited knowledge, incomplete characterization, or model approximations \cite{kiureghian2009aleatory}. While the former is irreducible, the latter can in principle be reduced through additional data, improved models, or refined identification procedures. Epistemic uncertainty is particularly essential in certification-oriented applications, where decisions must remain statistically justified even when data and computational resources are scarce.

Several recent contributions have emphasized the importance of explicitly propagating epistemic uncertainty in engineering problems. Methods based on imprecise probabilities have been proposed to handle severe knowledge gaps in physical models \cite{beer2013overview}, while non-intrusive Bayesian propagation frameworks, such as those for model calibration, have been developed to incorporate surrogate-model uncertainty and model discrepancy into the inference process \cite{kennedy2001bayesian}. Furthermore, likelihood-based approaches and other Bayesian updating techniques have been introduced to quantify and reduce epistemic uncertainty arising from sparse or interval data in complex systems \cite{sankararaman2013likelihood}. These advances in engineering find a parallel in modern machine learning, where methodological perspectives also  distinguish between epistemic and aleatory uncertainties, highlighting the need for explicit treatment of approximation error and limited data \cite{hullermeier2021aleatoric}. Despite these developments, many industrial workflows traditionally rely on established safety factors \cite{Hirsch2007,Eckert2019} to ensure structural integrity. While these margins provide a baseline for certification, the proposed approach aims to provide additional statistical transparency by explicitly quantifying epistemic uncertainties.

In aerospace certification, A-basis and B-basis values are conservative quantile estimates used to define allowable properties of composite materials. While historically established from extensive experimental testing datasets as standardized in the Composite Materials Handbook (CMH-17) \cite{MILHDBK17}, these are mandated by certification authorities (EASA CS-25.613 \cite{EASA_CS25}) to ensure that material and structural properties meet stringent statistical confidence requirements. The present work addresses the specific challenges of estimating these values within the framework of numerical modeling.  They are defined as:
\begin{itemize}
    \item The \textbf{A-basis value} is a tolerance bound associated with the 1st percentile, guaranteeing that at least 99\% of the population satisfies the performance requirement with 95\% confidence.
    \item The \textbf{B-basis value} is a tolerance bound associated with the 10th percentile, guaranteeing that at least 90\% of the population satisfies the performance requirement with 95\% confidence.
\end{itemize}
These quantities are fundamental to certification because they encapsulate both aleatory variability (e.g., material heterogeneity) and epistemic uncertainty (e.g., limited sample size, imperfect surrogate model, or uncertain input characterization). Their estimation is particularly challenging: composite testing campaigns are costly and time-consuming \cite{Lee2019}, and numerical simulations, though increasingly relied upon, are computationally expensive. To reduce cost, engineers turn to surrogate models, which introduce additional epistemic uncertainty from approximation error.


In this work, we investigate the impact of three major sources of epistemic uncertainty on the estimation of conservative quantiles within a probabilistic framework: probabilistic model identification, statistical estimation due to finite sampling, and surrogate-model approximation error. The proposed methodology offers a numerical alternative for estimating A-basis and B-basis, paving the way for design practices that account for uncertainty and are oriented towards certification.

This general numerical framework accounts explicitly for the effects of limited experimental data, surrogate-model approximation, and restricted evaluations of the high-fidelity model. It provides statistical confidence bounds incorporating all relevant sources of uncertainty. Beyond improving precision, the method has clear industrial relevance: it reduces reliance on safety factors defined by aerospace certification standards (CS-25) \cite{EASA_CS25} and delivers transparent, statistically justified numerical quantile bounds.

This paper is organized as follows. Section~\ref{section2} reviews relevant literature on variance reduction and conservative quantile estimation. Section~\ref{section3} formalizes the problem and details the sources of epistemic uncertainty. Section~\ref{section4} introduces the proposed quantile estimator, integrating importance sampling and control variates. Section~\ref{section5} applies the estimator to compute A- and B-basis values and performs a sensitivity analysis of each epistemic source. Section~\ref{section6} illustrates the method’s performance on numerical test cases and on an industrial-scale structural application, demonstrating its practical relevance for certification workflows. Section~\ref{conclusion} summarizes the contributions and outlines future research directions.

\section{Quantile estimators}
\label{section2}
In this section, we review classical estimators of a quantile $q_\alpha$ and its lower confidence bound, which form the foundation of the methodology proposed in this work. These estimators are introduced under the conventional assumption that the input probability distribution is perfectly known and that sufficiently large Monte Carlo samples can be drawn—assumptions that will be relaxed later to account for epistemic uncertainties. 

Let  \(\mathbf{X} = (X_1, \dots, X_d)^\top\) be a random input vector with probability density (pdf) $f_\mathbf{X}$, and let  \(Y = \phi(\mathbf{X})\) denote the model output with unknown pdf \(f_Y\) and cumulative distribution function (cdf) \(F_Y\). The \(\alpha\)-quantile (\(\alpha \in [0, 1]
 ) \)  $q_\alpha$ of \(Y\) is defined as
\begin{equation}
    q_\alpha = \inf \left\{ q \in \mathbb{R} \ \middle| \ F_Y(q) \geq \alpha \right\}.
\end{equation}

Most of the time, it is necessary to estimate the cdf of Y to provide a quantile estimator $\widehat{q}_\alpha$ of $q_\alpha$; and in most cases, a confidence interval (CI) at confidence level $\beta$ can be derived with:
\begin{equation}
    \left[
        \widehat{q}_{\alpha} - z_{(1+\beta)/2} \sqrt{{\mathbb{V}}(\widehat{q}_\alpha)},
        \ \widehat{q}_{\alpha} + z_{(1+\beta)/2}  \sqrt{{\mathbb{V}}(\widehat{q}_\alpha)}
    \right],
    \label{eq:quantile_ci_beta}
\end{equation}
where \( z_{(1+\beta)/2} \) denotes the $(1+\beta)/2$-quantile of the standard normal distribution and $\mathbb{V}$ denotes the variance under the simulation distribution. This CI provides the basis for constructing quantile conservative bounds. The choice between the lower and upper confidence bounds as the conservative estimate depends on the direction of the safety margin: lower bounds are retained for quantities to be minimized (e.g., strength), whereas upper bounds are used for quantities to be constrained from above. The A-basis and B-basis values are then obtained as the lower or the upper bound of a CI for respectively $\alpha=0.01$ and $\alpha=0.1$ with $\beta=0.95$.
 
In the literature, several statistical approaches have been proposed to estimate a quantile. In the following, we review the most commonly used estimators, which form the basis for constructing conservative quantile estimates.
\subsection{Empirical quantile estimator}
Let $\{ \mathbf{X}_i \}_{i=1}^{N_{\text{sim}}}$ be an i.i.d.(independent and identically distributed) input sample of size $N_{\text{sim}}$ with \(\mathbf{X}_i \sim f_\mathbf{X} \) and $\{Y_i = \phi(\mathbf{X}_i)\}_{i=1}^{N_{\text{sim}}}$  be the corresponding output sample.  
The cdf of $F_Y$ can be estimated by the empirical cdf $\widehat{F^{\text{MC}}_Y}$ with Monte Carlo methods such as \cite{DongNakayama2016}:
\begin{equation}
    \widehat{F^{\text{MC}}_Y}(y) = \frac{1}{N_{\text{sim}}} \sum_{i=1}^{N_{\text{sim}}} \mathbf{1}_{\phi(\mathbf{X}_i) \leq y},~ \mathbf{X}_i \sim f_\mathbf{X}. 
\end{equation}
The empirical $\alpha$-quantile estimate $\widehat{q_{\alpha}^{\text{MC}}}$ is the generalized inverse cdf:
\begin{equation}
    \widehat{q_{\alpha}^{\text{MC}}} = \inf \{ q \in \mathbb{R} \mid \widehat{F^{\text{MC}}_Y}(q) \geq \alpha \}.
\end{equation}
Equivalently, $\widehat{q_{\alpha}^{\text{MC}}}$ corresponds to the $\lceil \alpha N_{\text{sim}} \rceil$-th order statistic $Y_{(\lceil \alpha N_{\text{sim}} \rceil)}$, where $Y_{(1)} \leq \cdots \leq Y_{(N_{\text{sim}})}$ are the sorted sample values of $\{Y_i\}_{i=1}^{N_{\text{sim}}}$ and $\lceil \cdot \rceil$ is the ceiling function. 
The variance of $\widehat{q_{\alpha}^{\text{MC}}}$ (see Section 2.3.3 of \cite{Serfling2009}) is then:
\[ \mathbb{V}[\widehat{q_{\alpha}^{\text{MC}}}] = \frac{\alpha (1-\alpha)}{N_{\text{sim}} f_Y(q_\alpha)^2},\] which only reflects the uncertainty due to a finite sample number $N_{\text{sim}}$. This approach may require a substantial number of simulations to achieve a given level of variance for $\widehat{q_{\alpha}^{\text{MC}}}$.

\subsection{Quantile estimation with Importance Sampling}

Importance Sampling is a classical variance reduction technique. Instead of sampling from the original distribution $f_\mathbf{X}$, samples are drawn from a proposal distribution $g$, and quantile estimates are corrected using importance weights. Let $g$ be a given proposal density and $\{ \mathbf{X}_i \}_{i=1}^{N_{\text{sim}}}$ be an i.i.d.\ sample with \(\mathbf{X}_i \sim g \). The IS-based empirical cdf $\widehat{F^{\text{IS}}_Y}$ of $Y$ is then \cite{Arouna2004}
\begin{equation}
    \widehat{F^{\text{IS}}_Y}(y) = \frac{1}{N_{\text{sim}}} \sum_{i=1}^{N_{\text{sim}}}  \mathbf{1}_{\phi(\mathbf{X}_i) \leq y}\lambda^{\text{IS}}(\mathbf{X}_i),~\mathbf{X}_i \sim g,
\end{equation}
with
\begin{equation}
    \lambda^{\text{IS}}(\cdot) = \frac{f_\mathbf{X}(\cdot)}{g(\cdot)}.
\end{equation}
The IS-based $\alpha$-quantile estimate $\widehat{q_{\alpha}^{\text{IS}}}$ is computed with the generalized inverse cdf:
\begin{equation}
    \widehat{q_{\alpha}^{\text{IS}}} = \inf \{ q \in \mathbb{R} \mid \widehat{F^{\text{IS}}_Y}(q) \geq \alpha \}.
\end{equation}
The variance of $\widehat{q_{\alpha}^{\text{IS}}}$ is then given by \cite{Glynn1996}:
\begin{equation}
    \mathbb{V}[\widehat{q_{\alpha}^{\text{IS}}}] = \frac{\mathbb{E}_g[\mathbf{1}_{\phi(\mathbf{X}) \leq q_\alpha} \, (\lambda^{\text{IS}}(\mathbf{X}))^2] - \alpha^2}{N_{\text{sim}} f_Y(q_\alpha)^2}.
\end{equation}
$\mathbb{E}$ is the mathematical expectation and choosing an appropriate proposal density $g$ is critical for variance reduction as poor choices of $g$ can even increase the variance of $\widehat{q_{\alpha}^{\text{IS}}}$.
 
\subsection{Quantile estimation using Control Variate}
Control Variate (CV) is another classical variance reduction  \cite{Cannamela2008}. We assume that a surrogate model  \(\mathcal{M}\) of the function \(\phi\) is available. The \(\alpha\)-quantile $z_\alpha$ of $\mathcal{M}(\mathbf{X})$ is also assumed to be known or easy to estimate with high precision. The output $\mathcal{M}(\mathbf{X})$ may then serve as a control variate for the variable $Y$. Let $\{ \mathbf{X}_i \}_{i=1}^{N_{\text{sim}}}$ be an i.i.d.\ sample of $\mathbf{X}$ with \(\mathbf{X}_i \sim f_\mathbf{X} \).  The CV-based empirical cdf of $Y$ is:
\begin{align*}
  \widehat{F^{\text{CV}}_Y}(y) &= \frac{1}{N_{\text{sim}}} \sum_{i=1}^{N_{\text{sim}}} \mathbf{1}_{\phi(\mathbf{X}_i)  \leq y}  -C(y) \left( \frac{1}{N_{\text{sim}}} \sum_{i=1}^{N_{\text{sim}}} \mathbf{1}_{\mathcal{M}(\mathbf{X}_i) \leq z_\alpha}  -\mathbb{E}_{f_\mathbf{X}}[\mathbf{1}_{\mathcal{M}(\mathbf{X}) \leq z_\alpha}]\right) \\
  &=
  \widehat{F^{\text{MC}}_Y}(y) - C(y) (\widehat{h} -\alpha)
\end{align*}
with $C$ a coefficient and where the control variate $\widehat{h}$ is defined as
$$
  \widehat{h} = \frac{1}{N_{\text{sim}}} \sum_{i=1}^{N_{\text{sim}}} \mathbf{1}_{\mathcal{M}(\mathbf{X}_i) \leq z_\alpha}, \mathbf{X}_i \sim f_\mathbf{X}. 
$$
By definition of $z_\alpha$, $\mathbb{E}_{f_\mathbf{X}}[\mathbf{1}_{\mathcal{M}(\mathbf{X}) \leq z_\alpha}]=\alpha$. The efficiency of the control variate approach strongly depends on the choice of the control variable. In practice, it should be highly correlated with the quantity of interest while remaining inexpensive to evaluate. In this context, the surrogate model output $\mathcal{M}(\mathbf{X})$ provides a natural control variate, if it is able to capture most of the variability of the true model $\phi(\mathbf{X})$ at negligible computational cost.
The optimal parameter $C(y)$ \cite{Cannamela2008} in term of variance reduction is defined by:
\[
 {C}(y) =\frac{\text{Cov}[\widehat{F^{\text{MC}}_Y}(y),\widehat{h}]}{\mathbb{V}_{f_\mathbf{X}}[\widehat{h}]},
\]
where $\text{Cov}$ denotes the covariance. This choice minimizes the variance of the corrected estimator and corresponds to the optimal least-squares coefficient, as established in \cite{Cannamela2008}. 
The CV-based empirical $\alpha$-quantile is, as before, the generalized inverse cdf:
\begin{equation}
    \widehat{q_{\alpha}^{\text{CV}}} = \inf \{ q \in \mathbb{R} \mid \widehat{F^{\text{CV}}_Y}(q) \geq \alpha \}.
\end{equation}
The variance of this estimator is  \cite{Cannamela2008}:
\begin{equation}
    \mathbb{V}_{f_\mathbf{X}}[\widehat{q_{\alpha}^{\text{CV}}}] = \frac{ \alpha(1- \alpha)}{N_{\text{sim}} f_Y(q_\alpha)^2} [1-\rho^2],
\end{equation}
where $\rho$ is the correlation coefficient between $\mathbf{1}_{\phi(\mathbf{X}) \leq q_{\alpha}}$ and $\mathbf{1}_{\mathcal{M}(\mathbf{X}) \leq z_{\alpha}}$. If the surrogate model 
$\mathcal{M}$ is a good approximation of $\phi$, the correlation coefficient $\rho$ will be close to one, leading to a strong reduction in variance. For poor correlation ($\rho\approx 0$), the estimator reverts to the empirical one, ensuring robustness. Control variates allow weighting the contribution of the surrogate model according to its quality, without introducing any risk. Unlike surrogate model-based substitution approaches, the CV technique preserves the unbiasedness of the quantile estimator, making it particularly attractive for certification contexts.

\subsection{Wilks' Quantile Estimator}
An alternative to the asymptotic CI of the previous sections is the Wilks' non-parametric quantile tolerance bound \cite{Wilks1941}, which provides a guaranteed lower bound on a quantile estimator \( \widehat{q}_\alpha \) without requiring knowledge of the density function \( f_Y \).
Let \( Y_{(1)} \leq Y_{(2)} \leq \dots \leq Y_{(N_{\text{sim}} )} \) denote the ordered statistics of an i.i.d. sample of size \( N_{\text{sim}} \). Wilks showed that the following bound holds:
\begin{equation}
    \mathbb{P}\left(Y_{(r)} \leq q_{\alpha}\right) = 1 - \sum_{k=0}^{r-1} \binom{N_{\text{sim}}}{k} \alpha^k (1 - \alpha)^{N_{\text{sim}}-k},
    \label{eq:wilks_bound}
\end{equation}
where \( r \in \{1, \dots, N_{\text{sim}}\} \) is the rank of the order statistic and \( Y_{(r)} \) is the associated empirical quantile. To guarantee that the quantile \( q_{\alpha} \) is exceeded with probability at least \( \beta \), one must choose the rank \( r \) such that:
\begin{equation}
    1 - \sum_{k=0}^{r-1} \binom{N_{\text{sim}}}{k} \alpha^k (1 - \alpha)^{N_{\text{sim}}-k}  \geq \beta.
\end{equation}

This formulation allows the construction of a one-sided non-parametric confidence bound, given by the order statistic $Y_{(r)}$ where \( r \) is the smallest integer such that the cumulative binomial probability above exceeds \( \beta \). Under Wilks’ formulation, the smallest observed value 
\( Y_{(1)} \) from a sample of size \( N = 299 \) serves as a 95\%-confidence lower bound for $q_{0.01}$.

These classical estimators provide the foundation for conservative quantile estimation under idealized conditions—namely, perfect knowledge of the input distribution and sufficiently large sampling. In realistic engineering settings, however, such conditions rarely hold: probabilistic models are identified from limited data, computational budgets are constrained, and surrogate models introduce additional approximation error. The next section formalizes these challenges and introduces a framework for quantile estimation under mixed epistemic uncertainties, paving the way for the derivation of statistically justified A- and B-basis values from stochastic numerical models.

\section{Uncertainties in statistical estimation}
\label{section3}
 The purpose of this section is to identify sources of epistemic uncertainty that affect the reliability of statistical estimators, with a particular focus on quantile estimates used in certification standards (e.g., A-basis and B-basis values). In contrast to aleatory uncertainty, which is associated with inherent variability or intrinsic randomness, epistemic uncertainty originates from limited knowledge and is, at least in principle, reducible through additional information or modeling improvements.

\subsection{Input density estimation}
In practice, the input probabilistic model \( f_{\mathbf{X}} \) is seldom known \cite{Surget2024, sarazin2021}, and its knowledge is assumed in this article to be restricted to a \( N _{\text{test}}\)-sample built from physical experiments. Whilst direct measurements of physical experiments are considered here, one can also resort to indirect experiment samples obtained from intermediate models. The complexity of these experiments can limit \( N _{\text{test}}\), thus inducing a small data context. A random \(N_{\text{test}}\)-experiment sample is hereafter denoted by \( \mathbf{\widetilde{X}}_{\text{obs}} = \{ \mathbf{X}_{\text{obs}}^{(j)} \}_{j=1}^{N_{\text{test}}} \) where
\(
\mathbf{X}_{\text{obs}}^{(j)} = \left(X_{\text{obs}^1}^{(j)}, \ldots, X_{\text{obs}^d}^{(j)} \right)^T \sim f_{\mathbf{X}}
\) obtained either from direct measurements or intermediate experimental models. An estimate 
\( \widehat{f}_{\mathbf{X} \mid \mathbf{\widetilde{X}}_{\text{obs}} }(\cdot \mid\mathbf{\widetilde{x}}_{\text{obs}}) \) is then obtained through either parametric \cite{VanderVaart1998} (maximum likelihood, method of moments) or non-parametric approaches (kernel density estimation) \cite{Huang2013}. Small datasets and possible dependencies between variables introduce epistemic uncertainty in the inferred distribution. 

A first source of epistemic uncertainty arises thus from the statistical inference process, which depends on the experimental sample size $N_{\text{test}}$.  
The quality of the inferred probabilistic model is strongly affected by the amount of available data, and a small sample size may lead to bad identification.  
Neglecting this inference-related uncertainty typically leads to overconfident and potentially non-conservative quantile estimates.

\subsection{Statistical estimation}

A second source of epistemic uncertainty arises from the statistical estimation process itself \cite{Surget2024, menz2020}. As discussed in Section~\ref{section2}, classical quantile estimators rely on a finite number of independent simulations \( N_{\text{sim}} \), which introduces sampling variability in the estimated quantile \( \widehat{q}_{\alpha} \) \cite{CasellaBerger2002}. Even when the input probability distribution \( f_{\mathbf{X}} \) is perfectly known, the estimator remains random due to the finite sampling of the input space.  

Under standard regularity conditions, the variance of the empirical quantile estimator scales as
\[
\mathbb{V}[\widehat{q}_{\alpha}] \propto \frac{1}{N_{\text{sim}} f_Y(q_\alpha)^2},
\]
where \( f_Y(q_\alpha) \) denotes the value of the output pdf at the true quantile (it is not known in practice and is not easy to estimate). This expression highlights that the estimation uncertainty decreases slowly with the number of simulations—specifically, at a rate proportional to \( N_{\text{sim}}^{-1/2} \). Consequently, obtaining a small uncertainty on the estimated quantile requires a very large number of simulations.

This sampling-induced uncertainty becomes particularly problematic when estimating low quantiles (e.g., \( \alpha = 0.01 \) or \( 0.1 \)) \cite{CaeiroGomes2008}, which correspond to tail events that are rarely observed in limited Monte Carlo samples. In such cases, the estimator variance increases sharply, leading to potentially unreliable confidence bounds.  

In industrial and certification contexts, the number of high-fidelity simulations is often severely constrained by computational cost. As a result, purely empirical approaches may be infeasible or yield overly uncertain results. To mitigate this limitation, variance reduction techniques such as importance sampling or control variates (see Section~\ref{section2}) can be employed to improve estimator precision under a fixed computational budget. However, even when such techniques are applied, the epistemic uncertainty due to finite sampling remains an intrinsic component of the overall uncertainty budget.

\subsection{The surrogate model and associated uncertainty} 
In many engineering applications, the evaluation of high-fidelity computational models (e.g., finite element or computational fluid dynamics simulations) is computationally demanding. To alleviate this burden, surrogate models (or meta models) are employed as efficient approximations of the true model response \cite{forrester2009engineering, santner2003design}. They are trained on a limited set of model evaluations, commonly referred to as the Design of Experiments (DoE):
\[
\{ \mathbf{\widetilde{x}}_{\text{DoE}}, \widetilde{\mathbf{y}}_{\text{DoE}} \} \subset \mathcal{X}^{N_\text{DoE}} \times \mathbb{R}^{N_\text{DoE}},
\]
where \( \widetilde{\mathbf{y}}_{\text{DoE}} = \phi(\mathbf{\widetilde{x}}_{\text{DoE}}) \). Once trained, the surrogate model ($\mathcal{M}$) provides rapid predictions at new input locations and enables large-scale uncertainty analyses.

However, this approximation introduces an additional source of epistemic uncertainty due to limited training data and model truncation. When surrogate models are used in uncertainty propagation or reliability estimation, neglecting this error may result in biased or overconfident predictions \cite{menz2020}. To explicitly represent this modeling uncertainty, we consider the surrogate model output as a stochastic process:
$\widehat{Y} = \mathcal{M}(\mathbf{X}, \omega),$ where \( \omega \) denotes a random element capturing the variability between possible surrogate model realizations. Probabilistic surrogates such as Gaussian process \cite{Sacks1989} naturally provide this uncertainty representation, whereas deterministic models (e.g., polynomial chaos or neural networks) require complementary techniques such as bootstrap \cite{Efron1994} or Bayesian inference \cite{Gal2016}.

In practice, several realizations of the surrogate model, denoted \( \{ \mathcal{M}_k(\cdot) \}_{k=1}^{N_t} \), can be generated and used to propagate input uncertainty through Monte Carlo simulations. This approach captures both aleatory variability in the inputs and epistemic variability due to the surrogate model approximation, leading to more reliable and interpretable results. Nonetheless, if the surrogate model uncertainty is underestimated, the overall estimator may lose its conservatism, motivating the need for methods that preserve robustness even when the surrogate is imperfect.

\section{Quantile estimation under epistemic uncertainty using combined CV and IS techniques}
\label{section4}
The objective of this section is to propose a unified quantile estimator that will be used to explicitly accounts for the three sources of epistemic uncertainty identified earlier: probabilistic model identification, finite Monte Carlo sampling, and surrogate model approximation. The approach combines two complementary techniques: Importance Sampling and Control Variates. IS enables separate handling of the first two uncertainty sources by sampling from an auxiliary distribution distinct from the one inferred from experimental data, while CV leverages the surrogate model to reduce estimator variance without introducing bias. 

In practice, IS and CV are jointly applied to estimate the cdf of the model output, from which the target quantile \( \widehat{q}_{\alpha} \) and its confidence bound are obtained.

\subsection{Estimation of the cumulative distribution function by CV and IS}

We assume that a surrogate model \( \mathcal{M} \) is available, for which the $\alpha$-quantile \( z_{\alpha} \) of \( \mathcal{M}(\mathbf{X}) \) can be accurately estimated. With $g$ an IS auxiliary distribution, the proposed unbiased cdf estimator coupling CV and IS (denoted CV-IS) reads:
\begin{equation}
  \widehat{F_{Y}^{\text{CV-IS}}}(y) = \widehat{F^{\text{IS}}_Y}(y) - C(y) (\widehat{h} - \mathbb{E}_g[\mathbf{1}_{\mathcal{M}(\mathbf{X}) \leq z_\alpha}\lambda^{\text{IS}}(\mathbf{X})]),
\end{equation}
with
\begin{equation}
  \widehat{h} = \frac{1}{N_{\text{sim}}} \sum_{j=1}^{N_{\text{sim}}} \mathbf{1}_{\mathcal{M}(\mathbf{X}_j) \leq z_\alpha}\lambda^{\text{IS}}(\mathbf{X}_j), \quad \mathbf{X}_j \sim g.
\end{equation}
The optimal coefficient minimizing estimator variance is \cite{Cannamela2008}:
\begin{equation}
C(y) = \frac{\text{Cov}_g[\widehat{F^{\text{IS}}_Y}(y),\widehat{h}]}{\mathbb{V}_g[\widehat{h}]},
\label{c(y)}
\end{equation}
and its empirical estimate is given by:
\begin{equation}
  \widehat{C}(y) = \frac{\sum_{j=1}^{N_{\text{sim}}} (\mathbf{1}_{\phi(\mathbf{X}_j) \leq y}\lambda^{\text{IS}} (\mathbf{X}_j)- \widehat{F^{\text{IS}}_Y}(y))(\mathbf{1}_{\mathcal{M}(\mathbf{X}_j) \leq z_\alpha}\lambda^{\text{IS}} (\mathbf{X}_j) - \widehat{h})}{\sum_{i=1}^{N_{\text{sim}}} (\mathbf{1}_{\mathcal{M}(\mathbf{X}_i) \leq z_\alpha}\lambda^{\text{IS}} (\mathbf{X}_i)- \widehat{h})^2}.
\end{equation}
As \( z_\alpha \) is the \( \alpha \)-quantile of \( \mathcal{M} \), it follows that
\[
\mathbb{E}_{f_{\mathbf{X}}}[\mathbf{1}_{\mathcal{M}(\mathbf{X}) \leq z_\alpha}] 
= \mathbb{E}_g[\mathbf{1}_{\mathcal{M}(\mathbf{X}) \leq z_\alpha}\lambda^{\mathrm{IS}}(\mathbf{X})]
= \alpha.
\]
Thus, the cdf estimator simplifies to:
\begin{equation}
  \widehat{F_{Y}^{\text{CV-IS}}}(y) = \widehat{F^{\text{IS}}_Y}(y) - \widehat{C}(y) (\widehat{h} - \alpha)
  = \sum_{j=1}^{N_{\text{sim}}} W_j \mathbf{1}_{\phi(\mathbf{X}_j) \leq y},
  \label{eq:CV-cdf}
\end{equation}
 with the weights $W_j$ defined as 
\begin{equation}
  W_j =\left[ \frac{1}{N_{\text{sim}}} + \frac{ (\widehat{h} -\alpha) (\widehat{h}- \mathbf{1}_{\mathcal{M}(\mathbf{X}_j) \leq z_\alpha}\lambda^{\text{IS}}(\mathbf{X}_j)  )}{\sum_{i=1}^{N_{\text{sim}}} (\mathbf{1}_{\mathcal{M}(\mathbf{X}_i) \leq z_\alpha}\lambda^{\text{IS}}(\mathbf{X}_i) - \widehat{h})^2} \right] \lambda^{\text{IS}} (\mathbf{X}_j) 
  \label{IS-corr}
\end{equation}
It can be shown that (See \ref{Weigh_IS})
\begin{equation}
\sum_{j=1}^{N_{\text{sim}}} W_j \xrightarrow[N _{\text{sim}}\to \infty] {}1
\end{equation}
Minor negative weights may occur in Eq.~(\ref{IS-corr}) due to numerical noise or poor estimation of \( \widehat{C}(y) \). In such rare cases, weights are reverted to their original IS values, which preserves estimator unbiasedness.

The variance of the CV-IS estimator becomes:
\begin{align}
        \mathbb{V}_g[\widehat{F_{Y}^{\text{CV-IS}}}(y)] &= \mathbb{V}_g[\widehat{F^{\text{IS}}_Y}(y)] \,[1 - \rho^2(y)] \\ 
        &= \frac{\mathbb{E}_g[\mathbf{1}_{\phi(\mathbf{X}) \leq y}\lambda^{\text{IS}}(\mathbf{X})^2]- F_Y(y)^2}{N_{\text{sim}}}[1 - \rho^2(y)].
\end{align}
Here, \( \rho(y) \) denotes the correlation coefficient between \(\mathbf{1}_{\phi \leq y}\lambda^{\text{IS}}\) and \(\mathbf{1}_{\mathcal{M} \leq z_\alpha}\lambda^{\text{IS}}\). 
For strong correlation (\( \rho(y) \approx \pm 1 \)), the variance reduction is substantial; for weak correlation, the correction has negligible effect but does not degrade the estimator.

\subsection{Quantile estimation by CV-IS}

Once the cdf is estimated using Eq.~(\ref{eq:CV-cdf}), the corresponding quantile estimator is defined as:
\begin{equation}
  \widehat{q^{\text{CV-IS}}_{\alpha}} = \inf \left\{ q \in \mathbb{R} \ \mid \ \widehat{F_{Y}^{\text{CV-IS}}}(q)\geq \alpha \right\}.
  \label{eq:CV-quantile}
\end{equation}
This estimator is asymptotically normal, with reduced variance:
\begin{align}
    \mathbb{V}_{g}[\widehat{q^{\text{CV-IS}}_{\alpha}}] &= 
\mathbb{V}_g[\widehat{q_{\alpha}^{\text{IS}}}] \,[1 - \rho^2(q_{\alpha})]
= \frac{\mathbb{E}_g[\mathbf{1}_{\phi(\mathbf{X}) \leq q_{\alpha}}\lambda^{\text{IS}}(\mathbf{X})^2]- \alpha^2}{N_{\text{sim}}f_Y^2(q_\alpha)}[1 - \rho^2(q_{\alpha})].
  \label{Quantile_PPt}
\end{align}

The term \( 1 - \rho^2(q_{\alpha}) \) quantifies the effectiveness of the control variate in reducing variance. 
A strong correlation between \( \phi \) and \( \mathcal{M} \) leads to significant variance reduction, while poor correlation merely restores the baseline IS performance without biasing the result. 

\medskip

\section{Estimation and propagation under three sources of epistemic uncertainties}
\label{section5}
The proposed methodology provides a unified framework to propagate three main sources of epistemic uncertainty identified in Section~\ref{section3}. Their joint propagation by using the proposed quantile estimator from Section \ref{section4} constitutes the core contribution of this study.  

To handle these uncertainties, we adopt a hierarchical framework that creates a two-fold probabilistic structure. We introduce an augmented space where epistemic sources—probabilistic model identification parameters, finite simulation samples, and surrogate realizations—are treated as random variables alongside the aleatory inputs. This leads to a nested estimation problem: first, we compute a conditional CV-IS estimator of the quantile for a fixed realization of the epistemic variables; second, we perform an outer-loop Monte Carlo simulation in the augmented space. This double-level propagation ensures that the variability induced by all sources of epistemic uncertainty is captured in the final confidence bounds.

\subsection{Conditional CV-IS quantile estimation for a given scenario}
We first consider a single scenario defined by a fixed realization of each epistemic element:
\begin{itemize}
  \item a probabilistic input model inferred from an experimental dataset of size \(N_{\text{test}}\), denoted \(\mathbf{\widetilde{x}}_{\text{obs}}\), from which a conditional input distribution \(\widehat{f}_{\mathbf{X} \mid \mathbf{\widetilde{X}}_{\text{obs}}}(\cdot \mid \mathbf{\widetilde{x}}_{\text{obs}})\) is estimated;
  \item a simulation sample of size \(N_{\text{sim}}\), denoted \(\mathbf{\widetilde{X}} = \{\mathbf{X}_1,\dots,\mathbf{X}_{N_{\text{sim}}}\}\), drawn i.i.d.\ from a known auxiliary distribution \(g\);
  \item a surrogate model realization \(M(\cdot)\) of the surrogate model \(\mathcal{M}\) is available, for which the quantile \(z_\alpha\) has been estimated with high precision.
\end{itemize}

Given these inputs, the conditional CV-IS cumulative distribution estimator is written as
\begin{equation}
  \widehat{F_{Y}^{\text{CV-IS}}}(y, M,\mathbf{\widetilde{x}}_{\text{obs}},\mathbf{\widetilde{X}}) 
  =\sum_{j=1}^{N_{\text{sim}}} \left[ \frac{1}{N_{\text{sim}}} + \frac{ (\widehat{h} -\alpha) (\widehat{h} - \mathbf{1}_{M(\mathbf{X}_j) \leq z_\alpha}\lambda^{\text{IS}}_{ \mathbf{\widetilde{x}}_{\text{obs}}}(\mathbf{X}_j)  )}{\sum_{i=1}^{N_{\text{sim}}} (\mathbf{1}_{M(\mathbf{X}_i) \leq z_\alpha}\lambda^{\text{IS}}_{ \mathbf{\widetilde{x}}_{\text{obs}}}(\mathbf{X}_i) - \widehat{h})^2} \right] \lambda^{\text{IS}}_{ \mathbf{\widetilde{x}}_{\text{obs}}}(\mathbf{X}_j) \mathbf{1}_{\phi(\mathbf{X}_j) \leq y},
  \label{cdf_cv}
\end{equation}
where 
\[
\lambda_{\mathbf{\widetilde{x}}_{\text{obs}}}^{\text{IS}}(\mathbf{X}) = \frac{\widehat{f}_{\mathbf{X} \mid \mathbf{\widetilde{X}}_{\text{obs}}}({\mathbf{X} \mid \mathbf{\widetilde{x}}}_{\text{obs}})}{g(\mathbf{X})}
\quad \text{and} \quad
\widehat{h} = \frac{1}{N_{\text{sim}}} \sum_{i=1}^{N_{\text{sim}}} \mathbf{1}_{M(\mathbf{X}_i) \leq z_\alpha}\lambda_{\mathbf{\widetilde{x}}_{\text{obs}}}^{\text{IS}}(\mathbf{X}_i).
\]
The associated quantile estimator is then
\begin{equation}
  \widehat{q_{\alpha}^{\text{CV-IS}}}(M,\mathbf{\widetilde{x}}_{\text{obs}},\mathbf{\widetilde{X}})
  =\inf \left\{ q \in \mathbb{R} \ \mid \ \widehat{F_{Y}^{\text{CV-IS}}}(q,M,\mathbf{\widetilde{x}}_{\text{obs}},\mathbf{\widetilde{X}})\geq \alpha \right\}.
  \label{cv-cond-q}
\end{equation}
This conditional estimator inherits the properties of the CV-IS estimator defined in Section~\ref{section4}:  
it remains unbiased and its variance satisfies
\begin{align*}       
\mathbb{V}_g[\widehat{q_{\alpha}^{\text{CV-IS}}}(M,\mathbf{\widetilde{x}}_{\text{obs}},\mathbf{\widetilde{X}})] 
        &= \frac{\mathbb{E}_g[\mathbf{1}_{\phi(\mathbf{X}) \leq q_{\alpha}}\lambda_{\mathbf{\widetilde{x}}_{\text{obs}}}^{\text{IS}}(\mathbf{X})^2]- \alpha^2}{N_{\text{sim}}f_Y^2(q_\alpha)}[1 - \rho^2(q_{\alpha},M, \mathbf{\widetilde{x}}_{\text{obs}})],
\end{align*}
where \(\rho(q_{\alpha}, M, \mathbf{\widetilde{x}}_{\text{obs}})\) denotes the correlation coefficient between \(\mathbf{1}_{ \phi(\cdot) \leq q_\alpha}\lambda_{\mathbf{\widetilde{x}}_{\text{obs}}}^{\text{IS}}(\cdot)\) and \(\mathbf{1}_{M(\cdot) \leq z_\alpha}\lambda_{\mathbf{\widetilde{x}}_{\text{obs}}}^{\text{IS}}(\cdot)\).

\subsection{Proposed estimation under three sources of epistemic uncertainty}

In this section, we estimate the distribution of the quantile obtained by the CV--IS estimator while accounting for the three epistemic uncertainty sources introduced in Section~\ref{section3}. To this end, we embed all uncertainties into an augmented space defined by
\[
(\mathcal{M}, \widetilde{\mathbf{X}}_{\mathrm{obs}}, \widetilde{\mathbf{X}})
\in \Omega_q \times \mathcal{X}_{\mathrm{obs}}^{\otimes} \times \mathcal{X}^{\otimes},
\]
where $\Omega_q$ denotes the set of surrogate model realizations, 
$\mathcal{X}_{\mathrm{obs}}^{\otimes}\subset\mathbb{R}^{d\otimes N_{\mathrm{test}}}$ a set of experimental observations of size $N_{\mathrm{test}}$, 
and $\mathcal{X}^{\otimes}\subset\mathbb{R}^{d\otimes N_{\mathrm{sim}}}$ a simulation sample of size $N_{\mathrm{sim}}$.

The quantile estimator proposed in Eq.~(\ref{cv-cond-q}) can then be interpreted as a deterministic mapping
\[
\widehat{q_{\alpha}^{\text{CV-IS}}}
= \Phi(\mathcal{M}, \widetilde{\mathbf{X}}_{\mathrm{obs}}, \widetilde{\mathbf{X}}),\quad (\mathcal{M}, \widetilde{\mathbf{X}}_{\mathrm{obs}}, \widetilde{\mathbf{X}}) \sim f_{\mathcal{M}, \widetilde{\mathbf{X}}_{\mathrm{obs}}, \widetilde{\mathbf{X}}},
\]
which takes the three random inputs and returns an estimated quantile.  
In this augmented probabilistic setting, the quantile itself becomes a random variable.

\medskip

To propagate the associated uncertainty, a probability distribution is assigned to each epistemic source:
\begin{itemize}
\item experimental uncertainty is represented by the bootstrap distribution from the reference experimental data $\widetilde{\mathbf{x}}_{\mathrm{obs}}$;
\item Monte Carlo uncertainty is obtained by bootstrap resampling of a reference simulation sample $\widetilde{\mathbf{x}}$ ;
\item surrogate model uncertainty is represented either by direct sampling of surrogate model realizations (or by bootstrap resampling of the DoE used to train it).
\end{itemize}

Drawing $N$ independent triplets
\[
(\mathcal{M}_i,\widetilde{\mathbf{X}}_{\mathrm{obs}}^{(i)},\widetilde{\mathbf{X}}^{(i)}),
\qquad i=1,\dots,N,
\]
and applying the CV--IS estimator to each of them yields the quantile realizations
\[
\widehat{q_{\alpha,(i)}^{\text{CV-IS}}}
= \Phi(\mathcal{M}_i,\widetilde{\mathbf{X}}_{\mathrm{obs}}^{(i)},\widetilde{\mathbf{X}}^{(i)}).
\]
The collection $\{\widehat{q_{\alpha,(i)}^{\text{CV-IS}}}\}_{i=1}^{N}$ forms a Monte Carlo approximation of the distribution of the quantile taking all the considered epistemic uncertainties.

A key advantage of the proposed framework is that all realizations 
$\widehat{q_{\alpha,(i)}^{\text{CV-IS}}}$ are obtained \emph{without any additional evaluations} of the high-fidelity model and \emph{without any additional tests} used to identify the input density. 
The entire uncertainty propagation relies solely on bootstrap resampling and surrogate model realizations, 
making the method particularly suitable for industrial certification contexts with stringent computational budgets. The number of calls to the high-fidelity black-box model remains fixed and limited to \( N_{\text{sim}} + N_{\text{DoE}}\).

With the empirical distribution of the quantile estimator at hand, one can directly compute the conservative tolerance bounds of interest: the lower confidence limits associated with the 1st and 10th percentiles respectively yield the A- and B-basis values, thus completing the certification-oriented quantile estimation process.

The estimated asymptotic CI for the CV-IS quantile estimator 
\({\widehat{q_{\alpha}^{\text{CV-IS}}}}\), at confidence level \( \beta \) is directly computed from the empirical quantiles of the sample $\{\widehat{q_{\alpha,(i)}^{\text{CV-IS}}}\}_{i=1}^{N}$:
\begin{equation}
    \left[
    {\widehat{q_{\alpha,\inf}^{\text{CV-IS}}}}(\beta), \;
    {\widehat{q_{\alpha,\sup}^{\text{CV-IS}}}}(\beta)
    \right]
    \label{IC},
\end{equation}
where \({\widehat{q_{\alpha,\inf}^{\text{CV-IS}}}}(\beta), \) and \( {\widehat{q_{\alpha,\sup}^{\text{CV-IS}}}}(\beta) \) are the empirical quantiles at levels \( (1- \beta)/2 \) and \(  (1+\beta)/2 \) of the sample $\{\widehat{q_{\alpha,(i)}^{\text{CV-IS}}}\}_{i=1}^{N}$.
The A-basis and B-basis values correspond directly to the lower confidence bound quantile estimator  for specific \(\alpha\) and \(\beta\) (\({\text{A-basis}=\widehat{q_{0.01,\inf}^{\text{CV-IS}}}}(0.95) \) and \({\text{B-basis}=\widehat{q_{0.1,\inf}^{\text{CV-IS}}}}(0.95)\)). The estimates obtained using this approach are highly consistent and explicitly account for all epistemic uncertainties. Algorithm~\ref{alg:Elton} illustrates how this CI can be obtained in practice.
\begin{algorithm}
\caption{Estimation of a robust CI of the estimated quantile, incorporating all sources of epistemic uncertainty}
\label{alg:Elton}
\begin{algorithmic}[1]
\Require Quantile order $\alpha$, level of confidence interval $\beta$, reference $N_{\text{test}}$-experiment sample $\mathbf{\widetilde{x}}_{\text{test}}$, reference $N_{\text{DoE}}$-simulation sample $\mathbf{\widetilde{x}}_{\text{DoE}}$ with \(
\{ \widetilde{\mathbf{y}}_{\text{DoE}} = \phi(\mathbf{\widetilde{x}}_{\text{DoE}}) \}
\), size of BS N.
\State Build the surrogate \(\mathcal{M}(\cdot \mid \mathbf{\widetilde{x}}_{\text{DoE}}, \widetilde{\mathbf{y}}_{\text{DoE}}   )\)
\For{$k = 1$ to $N$}
    \State Resample $\widetilde{\mathbf{X}}_{\mathrm{obs}}^{(k)}$ from $\mathbf{\widetilde{x}}_{\text{test}}$
    \State Identify \(\widehat{f}_{\mathbf{X} \mid \mathbf{\widetilde{X}}_{\text{obs}}}(\cdot \mid \widetilde{\mathbf{X}}_{\mathrm{obs}}^{(k)}) \) from $\widetilde{\mathbf{X}}_{\mathrm{obs}}^{(k)}$
\EndFor
\If{the simulation samples $\mathbf{\widetilde{x}}$ have already been generated}
    \State $g \gets \text{density used to generate the simulation samples}$
\Else
   \State Build
    \(
        g(\cdot) = \frac{1}{N} \sum_{i = 1}^{N} \widehat{f}_{\mathbf{X} \mid \mathbf{\widetilde{X}}_{\text{obs}}}(\cdot \mid \widetilde{\mathbf{X}}_{\mathrm{obs}}^{(i)})\) from Eq. (\ref{eq:mixture_pdf2})
    \State Sample the set \(
    \mathbf{\widetilde{x}} = \left\{ \mathbf{x}_j, \quad j \in \llbracket 1, N_{\text{sim}} \rrbracket \right\} \) from $g$
    \State Evaluate \(\phi(\mathbf{x}_j)\) for \( j \in \llbracket 1, N_{\text{sim}} \rrbracket\) 
\EndIf

\For{$k = 1$ to $N$}
    \State Resample $\widetilde{\mathbf{X}}_{}^{(k)}$ from $\mathbf{\widetilde{x}}$
    \State Generate a surrogate model trajectory $\mathcal{M}_k$ from \(\mathcal{M}\)
    \State Estimate \(\widehat{q_{\alpha}^{\text{CV-IS}}}(\mathcal{M}_k, \widetilde{\mathbf{X}}_{\mathrm{obs}}^{(k)}, \widetilde{\mathbf{X}}_{}^{(k)})\) from Eq. (\ref{cv-cond-q})
\EndFor
\State Calculate the CI \(\left[
    {\widehat{q_{\alpha,\inf}^{\text{CV-IS}}}}(\beta), \;
    {\widehat{q_{\alpha,\sup}^{\text{CV-IS}}}}(\beta)
    \right] \) from Eq. (\ref{IC})
\State \Return $\left[
    {\widehat{q_{\alpha,\inf}^{\text{CV-IS}}}}(\beta), \;
    {\widehat{q_{\alpha,\sup}^{\text{CV-IS}}}}(\beta)
    \right]$

\end{algorithmic}
\end{algorithm}

\subsection{Importance Sampling auxiliary distribution}
\label{sec:IS_auxiliary_distribution}
There is flexibility in defining the auxiliary probability density function used within our methodology.  

When the simulation samples have already been generated according to a known sampling density, the auxiliary distribution \( g \) simply coincides with the pdf used to produce these simulation samples.  

Conversely, when the simulation samples have not yet been generated, the auxiliary pdf must be explicitly defined.  
In this case, we propose to use a mixture of target pdfs, interpreted as the average density over the identified models:
\begin{equation}
 \mathbf{x} \in \mathbb{R}^d, \quad
    g(\mathbf{x}) = \frac{1}{N} \sum_{i = 1}^{N} \widehat{f}_{\mathbf{X} \mid \mathbf{\widetilde{X}}_{\text{obs}}}(\mathbf{x} \mid \widetilde{\mathbf{X}}_{\mathrm{obs}}^{(i)})
    \label{eq:mixture_pdf2}.
\end{equation}
As suggested in \cite{OwenZhou2000}, this conservative strategy, which incorporates all identified conditional densities into the mixture, helps to mitigate numerical instabilities from IS. However, it is important to emphasize that constructing the auxiliary density \( g \) from the set of estimated conditional densities \( \{ \widehat{f}_{\mathbf{X} \mid \mathbf{\widetilde{X}}_{\text{obs}}}(\cdot\mid \widetilde{\mathbf{X}}_{\mathrm{obs}}^{(i)}) \}_{i=1}^{N}  \) does not violate the assumption of mutual independence between the random simulation sample \( \mathbf{\widetilde{X}} \) and the random observation sample \( \mathbf{\widetilde{X}}_{\text{obs}} \), as this independence assumption was made regardless of the specific choice of the density \( g \). Consequently, the independence remains valid even when employing the mixture density defined in Eq.~\eqref{eq:mixture_pdf2}. 

Finally, it is worth noting that, once chosen, the auxiliary density \( g \) remains fixed throughout the estimation process in the proposed framework.

\subsection{Variance decomposition and sensitivity indices (Sobol)}
In order to assess the contributions of each sources of epistemic uncertainties $\mathcal{M}$, $\mathbf{\widetilde{X}}_{\text{obs}}$ and $\mathbf{\widetilde{X}}$ on the variance of ${\widehat{q_{\alpha}^{\text{CV-IS}}}}(\mathcal{M},\mathbf{\widetilde{X}}_{\text{obs}},\mathbf{\widetilde{X}})$ separately, we can refer to the variance decomposition expression which is a classical tool in sensitivity analysis. In the augmented space, the variables \((\mathcal{M}, \mathbf{\widetilde{X}}_{\text{obs}}, \mathbf{\widetilde{X}})\) are independent. This assumption allows for the application of a standard Sobol sensitivity analysis  \cite{Sobol1993} in order to quantify the contribution of each input variable to the total variance. Such an analysis helps identify which source of epistemic uncertainty has the most significant impact on the estimator's variability. The first order Sobol indices $S$ and the total sobol indices $S^T$ are defined as:

\begin{align*}
  S_{\mathbf{\widetilde{X}}} &= \frac{\mathbb{V} \left[ \mathbb{E} [{\widehat{q_{\alpha}^{\text{CV-IS}}}}\mid \mathbf{\widetilde{X}}] \right]}{\mathbb{V}[{\widehat{q_{\alpha}^{\text{CV-IS}}}}]},
  \quad S_{\mathbf{\widetilde{X}}}^T = 1 - \frac{\mathbb{V} \left[ \mathbb{E} [{\widehat{q_{\alpha}^{\text{CV-IS}}}}\mid \mathcal{M},{\mathbf{\widetilde{X}}_{\text{obs}}} ] \right]}{\mathbb{V}[{\widehat{q_{\alpha}^{\text{CV-IS}}}}]}\\
  S_{{ \mathbf{\widetilde{X}}_{\text{obs}}}} &= \frac{\mathbb{V} \left[ \mathbb{E} [{\widehat{q_{\alpha}^{\text{CV-IS}}}}\mid { \mathbf{\widetilde{X}}_{\text{obs}}}] \right]}{\mathbb{V}[{\widehat{q_{\alpha}^{\text{CV-IS}}}}]}, 
  \quad 
  S_{\mathbf{\widetilde{X}}_{\text{obs}}}^T = 1 - \frac{\mathbb{V} \left[ \mathbb{E} [{\widehat{q_{\alpha}^{\text{CV-IS}}}}\mid \mathcal{M},\mathbf{\widetilde{X}} ] \right]}{\mathbb{V}[{\widehat{q_{\alpha}^{\text{CV-IS}}}}]} \\
  S_{\mathcal{M}} &= \frac{\mathbb{V} \left[ \mathbb{E} [{\widehat{q_{\alpha}^{\text{CV-IS}}}}\mid \mathcal{M}] \right]}{\mathbb{V}[{\widehat{q_{\alpha}^{\text{CV-IS}}}}]}, 
  \quad S_{\mathcal{M}}^T = 1 - \frac{\mathbb{V} \left[ \mathbb{E} [{\widehat{q_{\alpha}^{\text{CV-IS}}}}\mid {\mathbf{\widetilde{X}}_{\text{obs}}},\mathbf{\widetilde{X}} ] \right]}{\mathbb{V}[{\widehat{q_{\alpha}^{\text{CV-IS}}}}]}.
\end{align*}

This decomposition provides direct insight into the behavior of the control variate technique. Specifically, the influence of the surrogate model $\mathcal{M}$ is intrinsically linked to the optimal control coefficient $C(y)$ defined in Eq.~(\ref{c(y)}). For a weakly trained surrogate model, $C(y)$ tends toward zero, thereby reducing the interaction of the surrogate model with the estimator; consequently, the associated sensitivity indices ($S_{\mathcal{M}}$ and $S_{\mathcal{M}}^T$) are expected to be negligible. Conversely, when the surrogate model is well trained, this coefficient is non-zero, which creates a strong interaction between the surrogate model and the classical estimator, allowing for variance reduction.

The different Sobol indices can be estimated with the Pick-Freeze~\cite{deCastro2015} method. This estimation is performed solely from resampled inputs and surrogate model realizations, without requiring any additional evaluations of the expensive simulator. This makes it possible to assess, at negligible computational cost, the relative importance of each epistemic source on the final quantile uncertainty.

\section{Numerical applications}
\label{section6}
The feasibility and relevance of the proposed approach are illustrated
through two complementary case studies. The first example concerns an analytical
benchmark — a cantilever beam — which enables controlled assessment
of the estimator’s behavior under various uncertainty configurations.
The second case study illustrates the application to a representative industrial assembly—the cargo door interfaces of a commercial aircraft currently under development. These are representative of certification practices for composite structures under epistemic uncertainty.
\subsection{Deflection of a cantilever beam}
\subsubsection{Description of the model}
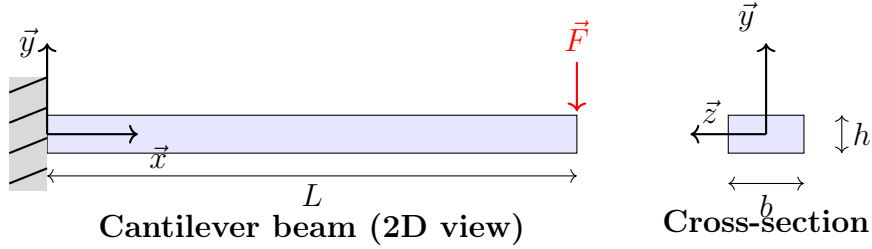
\begin{figure}[H]
\centering
\begin{tikzpicture}

\begin{scope}[shift={(0,0)}]

  \draw[fill=blue!10] (0,0) rectangle (7,0.5);

  \fill[gray!30] (-0.5,-0.5) rectangle (0,1);
  \foreach \y in {-0.4,0,0.4,0.8} {
    \draw[thick] (-0.5,\y) -- (0,\y+0.2);
  }

  \draw[->, thick, red] (7,1.2) -- (7,0.55);
  \node[above, red] at (7,1.2) {$\vec{F}$};

  \draw[<->] (0,-0.3) -- (7,-0.3);
  \node at (3.5,-0.55) {$L$};

  \node at (3.5,-1.0) {\textbf{Cantilever beam (2D view)}};

  \coordinate (O) at (0,0.25);
  \draw[->, thick] (O) -- ++(1.2,0) node[below right] {$\vec{x}$};
  \draw[->, thick] (O) -- ++(0,1.2) node[left] {$\vec{y}$};
\end{scope}

\begin{scope}[shift={(9,0)}]

  \draw[fill=blue!10] (0,0) rectangle (1,0.5);

  \draw[<->] (1.5,0) -- (1.5,0.5);
  \node[right] at (1.5,0.25) {$h$};

  \draw[<->] (0,-0.4) -- (1,-0.4);
  \node[below] at (0.5,-0.4) {$b$};

  \node at (0.5,-0.9) {\textbf{Cross-section}};
  \coordinate (O) at (0.5,0.25);
  \draw[->, thick] (O) -- ++(0,1.2) node[above left] {$\vec{y}$};
  \draw[->, thick] (O) -- ++(-1,0) node[above right] {$\vec{z}$};
\end{scope}

\end{tikzpicture}

\caption{Illustration of a cantilever beam of length $L$, Young’s modulus $E_{\text{YM}}$ and cross-section $b \times h$ with an applied transverse load $F$.}
\label{fig:cantilever_beam}
\end{figure}

\begin{figure}[H]
    \centering
    \includegraphics[trim={0 10cm 0 10.8cm},clip,scale=0.35]{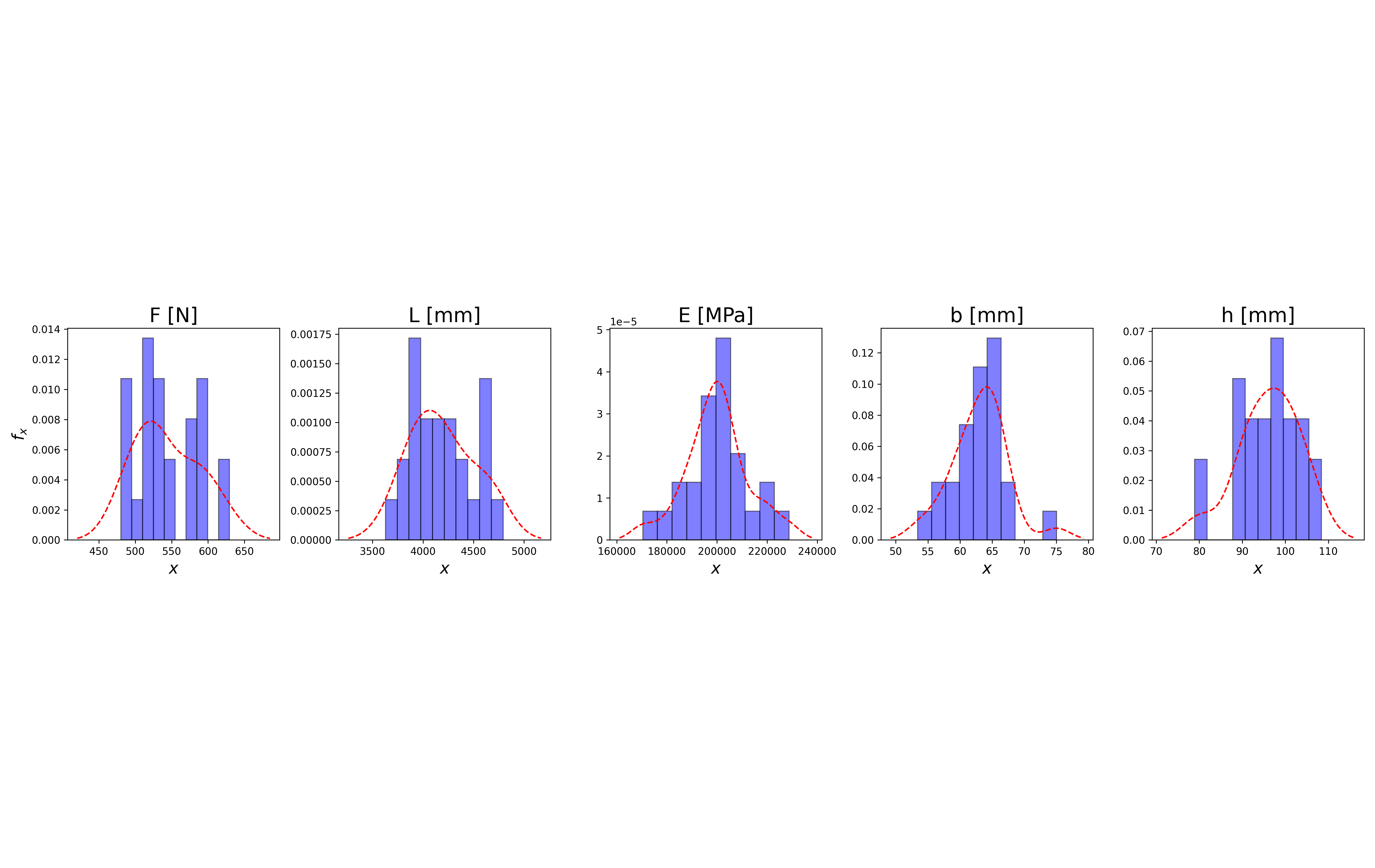}     
\caption{Distribution and Kernel Density Estimation (KDE) of the test sample for the Cantilever Beam inputs.}
\label{Test_sample}
\end{figure}

Consider a cantilever beam test case where the mean deflection at the free end is induced by the applied load represented in Figure \ref{fig:cantilever_beam}. The function $\phi$ writes in the following analytical form with d = 5 uncertain inputs:
\begin{equation}
  \phi(F,L,E_{YM},b,h)= \frac{4FL^3}{E_{YM}bh^3},
\end{equation}
where $F$ is the transverse load applied at the free end, $L$ is the beam length and $E_{YM}$ is the Young’s modulus. The parameters $b$ and $h$ represent the width and height of the rectangular cross-section, respectively. The actual distribution of each input variable is listed in Table \ref{tab:distributions_cantilever} from \cite{li2017reliability} and all the variables are assumed independent. However, this probabilistic model is assumed to be unknown and it only enables the generation of the $N_{\text{test}}$-experiment sample \(\mathbf{\widetilde{x}}_{\text{test}}\) on which a KDE (kernel density estimation) is performed (See Figure \ref{Test_sample}).
\begin{table}[H]
\centering
\caption{Input random variable distributions for the cantilever beam test case. CV denotes the coefficient of variation.}
\label{tab:distributions_cantilever}
\vspace{0.5em}
\begin{tabular}{@{}llccc@{}}
\toprule
\textbf{Variable} & \textbf{Symbol, Units} & \textbf{Distribution} & \textbf{Mean} & \textbf{CV} \\
\midrule
Transverse load   & $F$, N             & Lognormal & 556.8  & 0.08 \\
Length            & $L$, mm            & Normal    & 4290   & 0.10 \\
Young’s modulus   & $E_{\text{YM}}$, MPa & Lognormal & 2.105  & 0.06 \\
Width             & $b$, mm            & Normal    & 62     & 0.10 \\
Height            & $h$, mm            & Normal    & 98.7   & 0.10 \\
\bottomrule
\end{tabular}
\end{table}

\subsubsection{Results}

The auxiliary density used in this case for simulation samples is: 
\begin{equation*}
 \mathbf{x} \in \mathbb{R}^d, \quad
    g(\mathbf{x}) = \frac{1}{N} \sum_{i = 1}^{N} \widehat{f}_{\mathbf{X} \mid \widetilde{\mathbf{X}}_{\text{obs}}}(\mathbf{x} \mid \widetilde{\mathbf{X}}_{\mathrm{obs}}^{(i)}).
\label{eq:mixture_pdf}
\end{equation*}
This density is used for the purpose of generating the reference simulation sample \(\widetilde{\mathbf{x}}\), composed of \(N_{\text{sim}}\) samples. The surrogate models employed are Gaussian processes~\cite{rasmussen2006gaussian}, for which the trajectories can be efficiently generated using the Karhunen–Loève (KL) decomposition~\cite{Villemonteix2009}. Furthermore, the BS sample size is fixed at \(N = 10^3\). The objective is to estimate both A-basis and B-basis under the presence of three sources of epistemic uncertainty.

The different A-basis and B-basis value estimates are obtained for different given values of \((N_{\text{test}}, N_{\text{DoE}}, N_{\text{sim}}) \). For each set of values, we also provide the corresponding Sobol sensitivity indices (first-order and total-order) obtained with PF method.
\begin{figure}[H]
    \centering
    \includegraphics[width=1.25\textwidth]{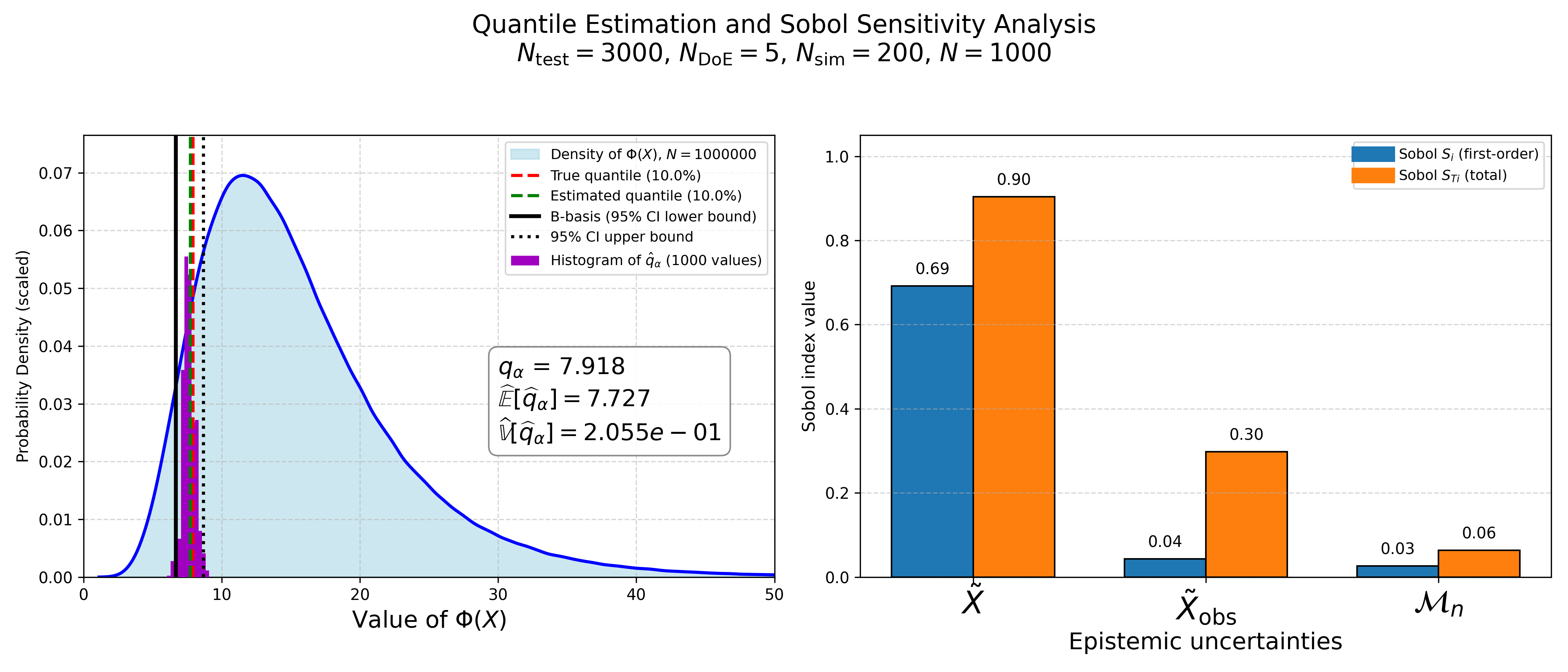}
\caption{Result for Monte Carlo sampling uncertainties only, good input identification and poor surrogate model.
\textbf{Left:} histogram of the estimated quantile and the corresponding B-basis; 
\textbf{Right:} first-order and total Sobol indices for all sources of epistemic uncertainty.}
    \label{result:fig1}
\end{figure}
Figure~\ref{result:fig1} was obtained with 
\((N_{\text{sim}}, N_{\text{test}}, N_{\text{DoE}}) = (200, 3000, 5)\), 
which ensures that there is sufficient data to learn the input distributions, 
while the surrogate model is poorly trained. On the left of the figure, we show the B-basis associated with this dataset 
and the variance of the quantile estimator. 
On the right, the sensitivity indices indicate that sampling uncertainty is predominant, 
as expected. The surrogate model sensitivity indices remain negligible, even when the surrogate model is poorly trained, an effect guaranteed by the use of control variates. Thus, a poorly trained surrogate model does not introduce bias or affect the estimator.

\begin{figure}[H]
    \centering
    \includegraphics[width=1.25\textwidth]{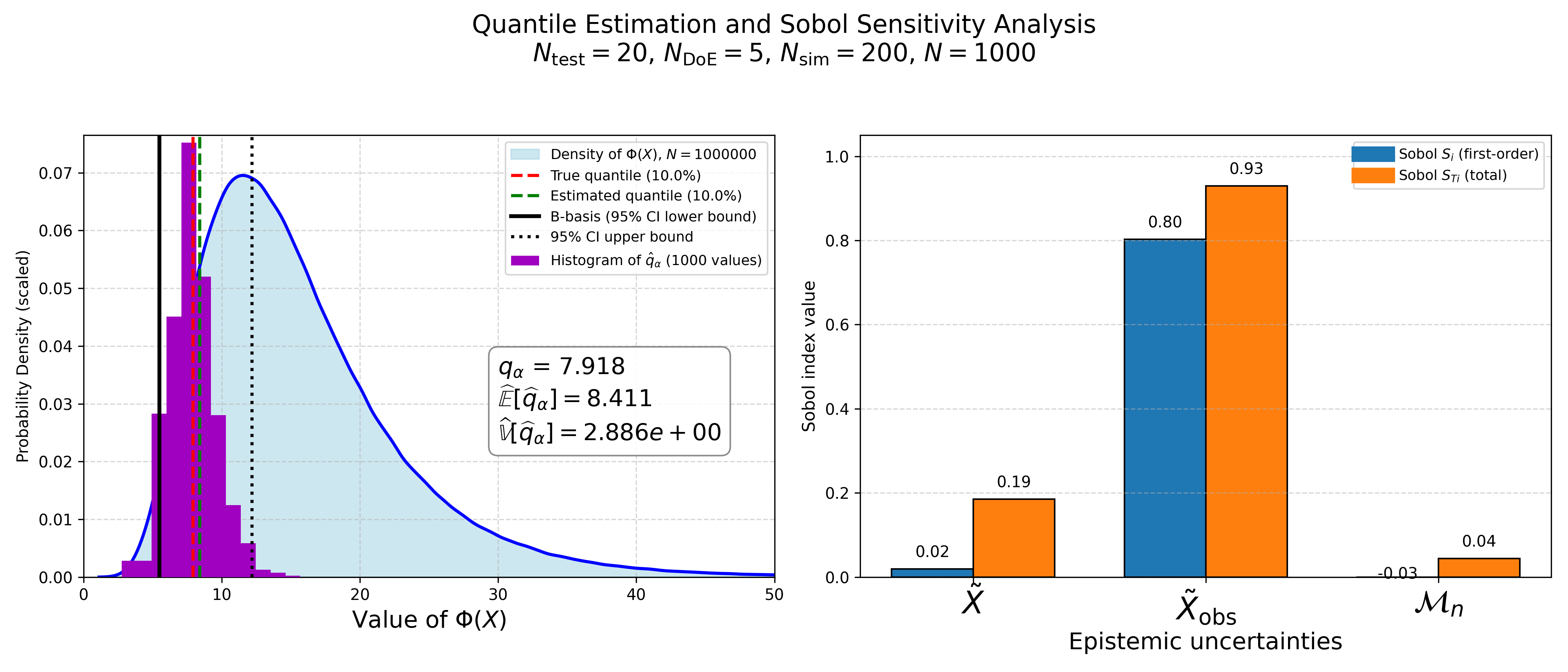}
\caption{Result for Monte Carlo sampling uncertainties, bad input identification and poor surrogate model. 
    \textbf{Left:} histogram of the estimated quantile and the corresponding B-basis; 
\textbf{Right:} first-order and total Sobol indices for all sources of epistemic uncertainty.}
    \label{result:fig2}
\end{figure}
Figure~\ref{result:fig2} was obtained with 
\((N_{\text{sim}}, N_{\text{test}}, N_{\text{DoE}}) = (200, 20, 5)\), which implies that there is insufficient data to accurately identify the input distributions (leading to high variability), while the surrogate model remains poorly trained. On the left of the figure, we present the B-basis associated with this dataset together with the variance of the quantile estimator (which has significantly increased (factor 14) compared to that in Figure \ref{result:fig1}. On the right, the sensitivity indices confirm that the variability related to the identification of input distributions is predominant, as expected. The sensitivity indices associated with the surrogate model remain negligible, even in the case of poor training, an effect ensured by the use of control variates. Therefore, a poorly trained surrogate model does not introduce bias nor does it affect the estimator.

\begin{figure}[H]
    \centering
    \includegraphics[width=1.25\textwidth]{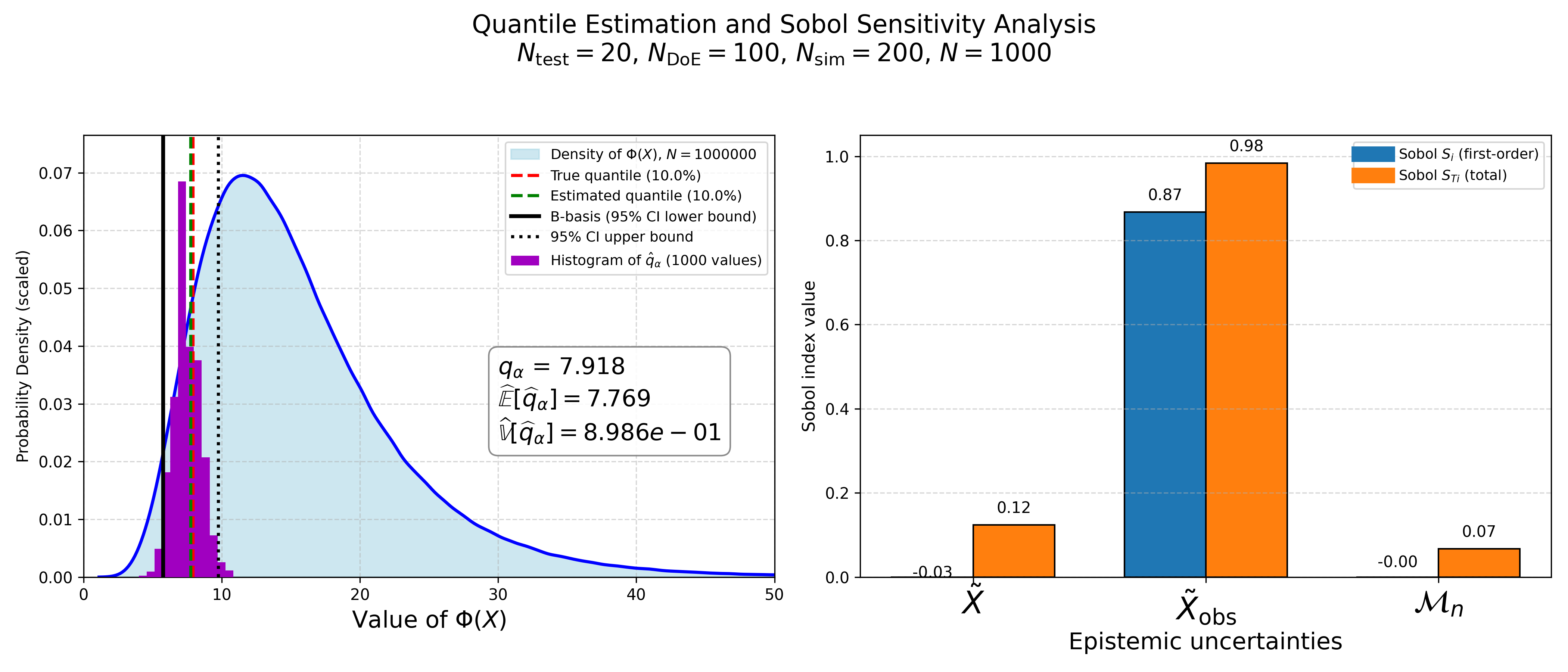}
\caption{Result for Monte Carlo sampling uncertainties, bad input identification and good surrogate model.
\textbf{Left:} histogram of the estimated quantile and the corresponding B-basis; 
\textbf{Right:} first-order and total Sobol indices for all sources of epistemic uncertainty.}
    \label{result:fig3}
\end{figure}
Figure~\ref{result:fig3} was obtained with 
\((N_{\text{sim}}, N_{\text{test}}, N_{\text{DoE}}) = (200, 20, 100)\), 
which ensures that there is insufficient data to accurately learn the input distributions (high variability). 
In this case, however, the surrogate model is well trained. The direct consequence is a reduction in the variance (factor 0.3)  of the quantile estimator, 
resulting in a more accurate estimation of the B-basis. The sensitivity analysis shows that the Sobol indices of the surrogate model remain low, which is expected, since the surrogate model helps the estimator converge more quickly without degrading its accuracy.


Table~\ref{tab:quantile_estimation1} details the A-basis estimation results for various combinations of \((N_{\text{sim}}, N_{\text{test}}, N_{\text{DoE}})\), while Table~\ref{tab:quantile_estimation2} presents the corresponding results for the B-basis.

\begin{table}[H]
\centering
\caption{Estimation of the \(\alpha\)-quantile with \(\alpha = 0.01\), \(\beta = 0.95\), and true quantile \(q_{\alpha} = 4.89\).
The reported statistics include the empirical mean and variance of the estimator $\widehat{q_{\alpha}^{\text{CV-IS}}}$, the CI$_{0.95}$ = $[\widehat{q_{\alpha,\inf}^{\text{CV-IS}}}, \widehat{q_{\alpha,\sup}^{\text{CV-IS}}}]$ at level $0.95$, and Sobol’ sensitivity indices (first-order $\widehat{S}$ and total $\widehat{S}^{T}$) computed for three sources of epistemic uncertainty. The lower bound $\widehat{q_{\alpha,\inf}^{\text{CV-IS}}}$ corresponds to the \textbf{A-basis} value.}
\begin{tabular}{
  c
  S[table-format=2.2]
  S[table-format=1.2]
  c
  c
}
\toprule
\((N_{\text{sim}}, N_{\text{test}}, N_{\text{DoE}})\) &
{\(\widehat{\mathbb{E}}[\widehat{q_{\alpha}^{\text{CV-IS}}}]\)} &
{\(\widehat{\mathbb{V}}[\widehat{q_{\alpha}^{\text{CV-IS}}}]\)} &
$[\text{A-basis}, \widehat{q_{\alpha,\sup}^{\text{CV-IS}}}]$ &
{\([\widehat{S}_{\mathbf{\widetilde{X}}}, \widehat{S}^T_{\mathbf{\widetilde{X}}}, \widehat{S}_{\mathbf{\widetilde{X}}_{\text{obs}}}, \widehat{S}^T_{\mathbf{\widetilde{X}}_{\text{obs}}}, \widehat{S}_{\mathcal{M}}, \widehat{S}^T_{\mathcal{M}}]\)} \\
\midrule
(100, 700, 3)    & 5.05 & 1.47 & [3.95, 7.16] & (0.50, 0.81, 0.21, 0.49, 0.01, 0.01) \\
(100, 700, 70)   & 4.87 & 0.89 & [3.04, 5.92] & (0.24, 0.89, 0.09, 0.78, 0.04, 0.20) \\
(300, 30, 3)     & 4.82 & 0.53 & [3.60, 5.80] & (0.21, 0.56, 0.38, 0.82, 0.03, 0.15) \\
(300, 30, 100)   & 4.55 & 0.45 & [3.59, 5.60] & (0.23, 0.75, 0.28, 0.82, 0.02, 0.26) \\
(300, 300, 3)    & 3.92 & 0.60 & [2.73, 5.29] & (0.20, 0.64, 0.33, 0.77, -0.01, 0.15) \\
(300, 300, 100)  & 4.22 & 0.37 & [3.20, 5.23] & (0.04, 0.84, 0.18, 0.98, 0.01, 0.21) \\
\bottomrule
\end{tabular}
\label{tab:quantile_estimation1}
\end{table}

\begin{table}[H]
\centering
\caption{Estimation of the $\alpha$-quantile with $\alpha = 0.10$, confidence level $\beta = 0.95$, and true quantile $q_{\alpha} = 7.89$. 
The reported statistics include the empirical mean and variance of the estimator $\widehat{q_{\alpha}^{\text{CV-IS}}}$, the CI$_{0.95}$ = $[\widehat{q_{\alpha,\inf}^{\text{CV-IS}}}, \widehat{q_{\alpha,\sup}^{\text{CV-IS}}}]$ at level $0.95$, and Sobol’ sensitivity indices (first-order $\widehat{S}$ and total $\widehat{S}^{T}$) computed for three sources of epistemic uncertainty. The lower bound $\widehat{q_{\alpha,\inf}^{\text{CV-IS}}}$ corresponds to the \textbf{B-basis} value.}
\begin{tabular}{
  c
  S[table-format=2.2]
  S[table-format=1.2]
  c
  c
}
\toprule
\((N_{\text{sim}}, N_{\text{test}}, N_{\text{DoE}})\) &
{\(\widehat{\mathbb{E}}[\widehat{q_{\alpha}^{\text{CV-IS}}}]\)} &
{\(\widehat{\mathbb{V}}[\widehat{q_{\alpha}^{\text{CV-IS}}}]\)} &
$[\text{B-basis}, \widehat{q_{\alpha,\sup}^{\text{CV-IS}}}]$  &
{\([\widehat{S}_{\mathbf{\widetilde{X}}}, \widehat{S}^T_{\mathbf{\widetilde{X}}}, \widehat{S}_{\mathbf{\widetilde{X}}_{\text{obs}}}, \widehat{S}^T_{\mathbf{\widetilde{X}}_{\text{obs}}}, \widehat{S}_{\mathcal{M}}, \widehat{S}^T_{\mathcal{M}}]\)}  \\
\midrule
(100, 700, 3)    & 7.32 & 0.35 & [6.17, 8.50] & (0.50, 0.85, 0.16, 0.41, 0.02, 0.14) \\
(100, 700, 70)   & 7.43 & 0.16 & [6.80, 8.27] & (0.02, 0.75, 0.24, 0.89, 0.03, 0.53) \\
(300, 30, 3)     & 7.58 & 1.70 & [4.78, 9.97] & (0.09, 0.25, 0.72, 0.93, 0.01, 0.06) \\
(300, 30, 100)   & 7.45 & 0.95 & [5.47, 9.42] & (0.01, 0.10, 0.90, 0.99, 0.03, 0.09) \\
(300, 300, 3)    & 7.70 & 0.40 & [6.60, 9.06] & (0.35, 0.67, 0.28, 0.63, -0.01, 0.07) \\
(300, 300, 100)  & 7.44 & 0.17 & [6.61, 8.15] & (0.01, 0.49, 0.55, 0.93, 0.00, 0.47) \\
\bottomrule
\end{tabular}
\label{tab:quantile_estimation2}
\end{table}

When examining these results, we observe that the CI always contains the true quantile, as all sources of epistemic uncertainty have been accounted for. The sensitivity indices indicate that, in general, the first-order indices of the surrogate model are nearly zero. This behavior is expected when control variates are employed, since they ensure that the surrogate model error does not deteriorate the final estimator.

Regarding the total-effect index of the surrogate model, one can note that a poorly trained surrogate model leads to a very small sensitivity index, while a well-trained surrogate model yields a larger one. This is again consistent with the fact that, for a weakly trained surrogate model, the control coefficient $C(y)$  tends toward zero, thereby reducing the interaction of the surrogate model with the estimator. Conversely, when the surrogate model is well trained, this coefficient is non-zero, which creates interaction between the surrogate model and the classical estimator, ultimately allowing variance reduction. This observation justifies the behavior of the total-effect sensitivity index of the surrogate model.

\subsubsection{Comparison with an idealized reference (Wilks' Method)}

To demonstrate the adaptive conservatism of the proposed framework, we compare our results against a benchmark based on the classical Wilks' method.
To ensure a fair comparison, the analysis is conducted at an equivalent computational budget.
To establish the \textbf{Wilks B-basis} (represented by red crosses), we performed a total of $N \times N_{\text{sim}}$ simulations sampled directly from the true probability density function. For each batch of size $N_{\text{sim}}$, we estimated a B-basis value using the standard Wilks formula; the final reference value corresponds to the average of these $N$ estimates.

\begin{figure}[H]
    \centering
    \includegraphics[width=1.0\textwidth]{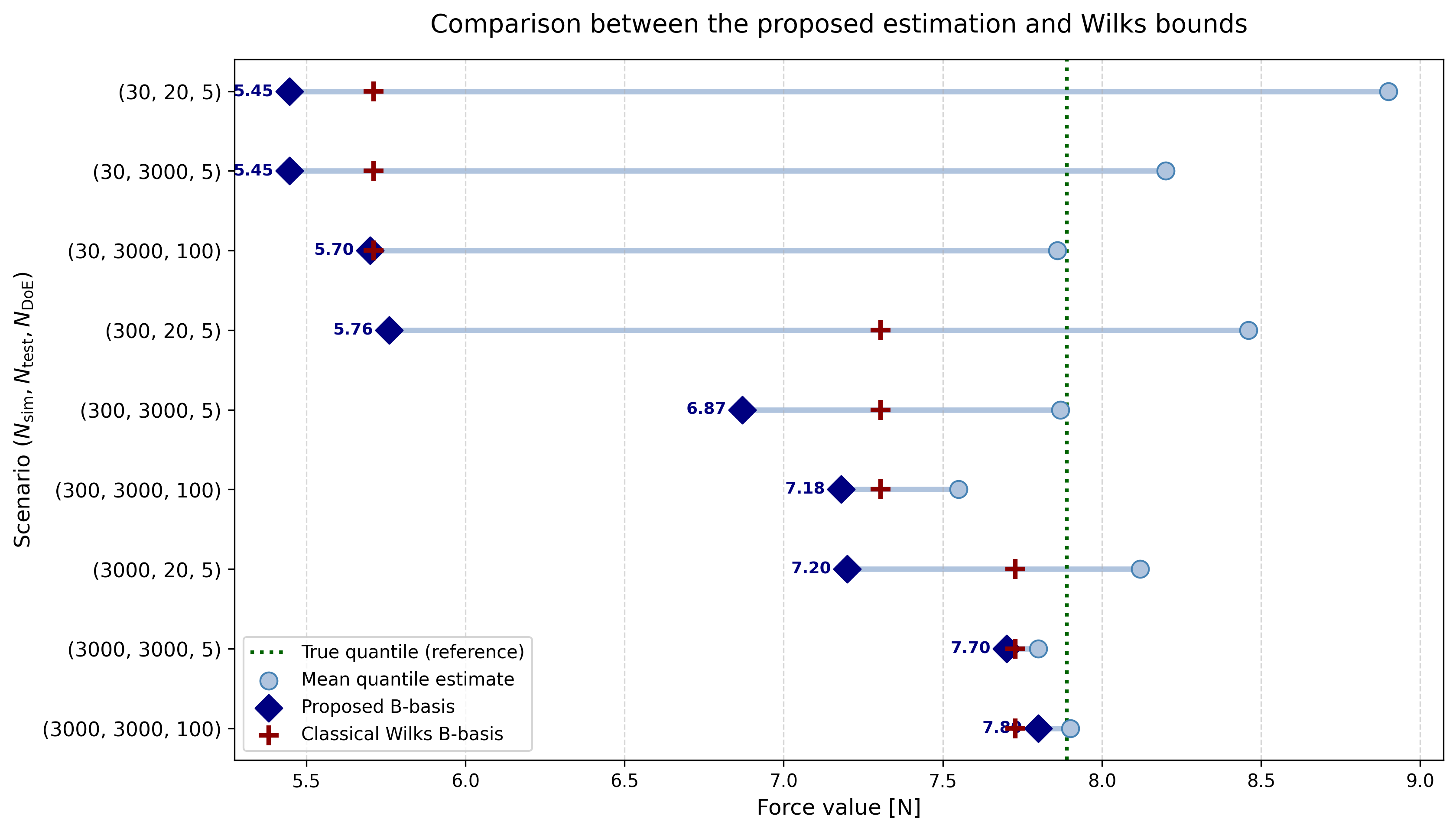} 
    \caption{Assessment of conservatism. The vertical green line represents the true quantile. The proposed B-basis (dark blue diamond) is compared to the average Wilks value (\textbf{red cross}).}
    \label{fig:wilks_comparison}
\end{figure}

Figure~\ref{fig:wilks_comparison} visualizes the results across different scenarios of data availability defined by the triplet $(N_{\text{sim}}, N_{\text{test}}, N_{\text{DoE}})$.
The vertical green dotted line represents the true theoretical quantile (Target).
For each scenario, the dumbbell plot connects the mean quantile estimate (light blue circle) to the proposed B-basis (dark blue diamond). The red crosses (+) indicate the averaged Wilks' reference.

This visualization highlights two key behaviors of the proposed CV-IS framework:

\begin{itemize}
    \item \textbf{Adaptive Conservatism in Data-Scarce Regimes:} in scenarios with low data availability (e.g., top rows: $N_{\text{sim}}=30$), the proposed B-basis (dark blue diamond) is significantly lower than the classical Wilks value (red cross). This is a desirable feature: unlike Wilks' method, which only accounts for sampling variability, our framework explicitly penalizes the quantile estimate to account for the epistemic uncertainty stemming from the surrogate model and the limited input data.
    
    \item \textbf{Convergence with Increased Information:} as the computational budget and experimental data increase (bottom rows: $N_{\text{sim}}=3000$), the gap between the mean estimate, the proposed B-basis, and the Wilks reference narrows. In the most data-rich scenario (3000, 3000, 100), the proposed method converges toward the true quantile (green line), demonstrating that the conservatism is relaxed as epistemic uncertainties are reduced.
\end{itemize}

\subsection{Stochastic analysis of the main deck cargo door interfaces of an Airbus commercial aircraft}
\subsubsection{Description of the model}
The second case study illustrates the application of our methodology to uncertainty management in the critical joints of the composite fuselage of a future \text{Airbus commercial aircraft}, focusing specifically on the interfaces of the main cargo door. The numerical model used is based on the Global Finite Element Method (\text{GFEM}), allowing an accurate representation of the structural complexity in this area. The studied interface comprises $m \in \mathbb{N}$ (\text{such that } $m > 8)$ critical attachment points, each characterized by a geometric gap that directly influences the force transmitted under aircraft loading. The objective of this study is to propagate the uncertainties associated with these interface gaps to quantify their impact on the transmitted loads at the interface (in a similar approach than one described in \cite{Capasso2023WCSMO}).

The input variables are the $m$ independent geometric gaps, represented as truncated normal distributions with bounds and means defined by tolerance analyses. As this future \text{Airbus commercial aircraft} was not yet in production, their variances were estimated from measurements on other aircraft at the Final Assembly Line, introducing epistemic uncertainty:
\[
\mathbf{X} = (X_1, \ldots, X_{m}), \quad X_i \sim \mathcal{N}_{[a_i,b_i]}(\mu_i, \sigma_i^2), \; i = 1, \ldots, m.
\]
The variance $\sigma_i^2$ for each gap is estimated empirically from the available measurements using the standard unbiased estimator:
\[
\widehat{\sigma_i^2} = \frac{1}{N_{\text{test}} - 1} \sum_{k=1}^{N_{\text{test}}} (x_{i,k} - \bar{x}_i)^2,
\]
where $x_{i,k}$ represents the $k$-th measurement of the $i$-th gap, and $\bar{x}_i$ is the sample mean.

\begin{figure}[H]
    \centering
    \includegraphics[width=1.2\textwidth]{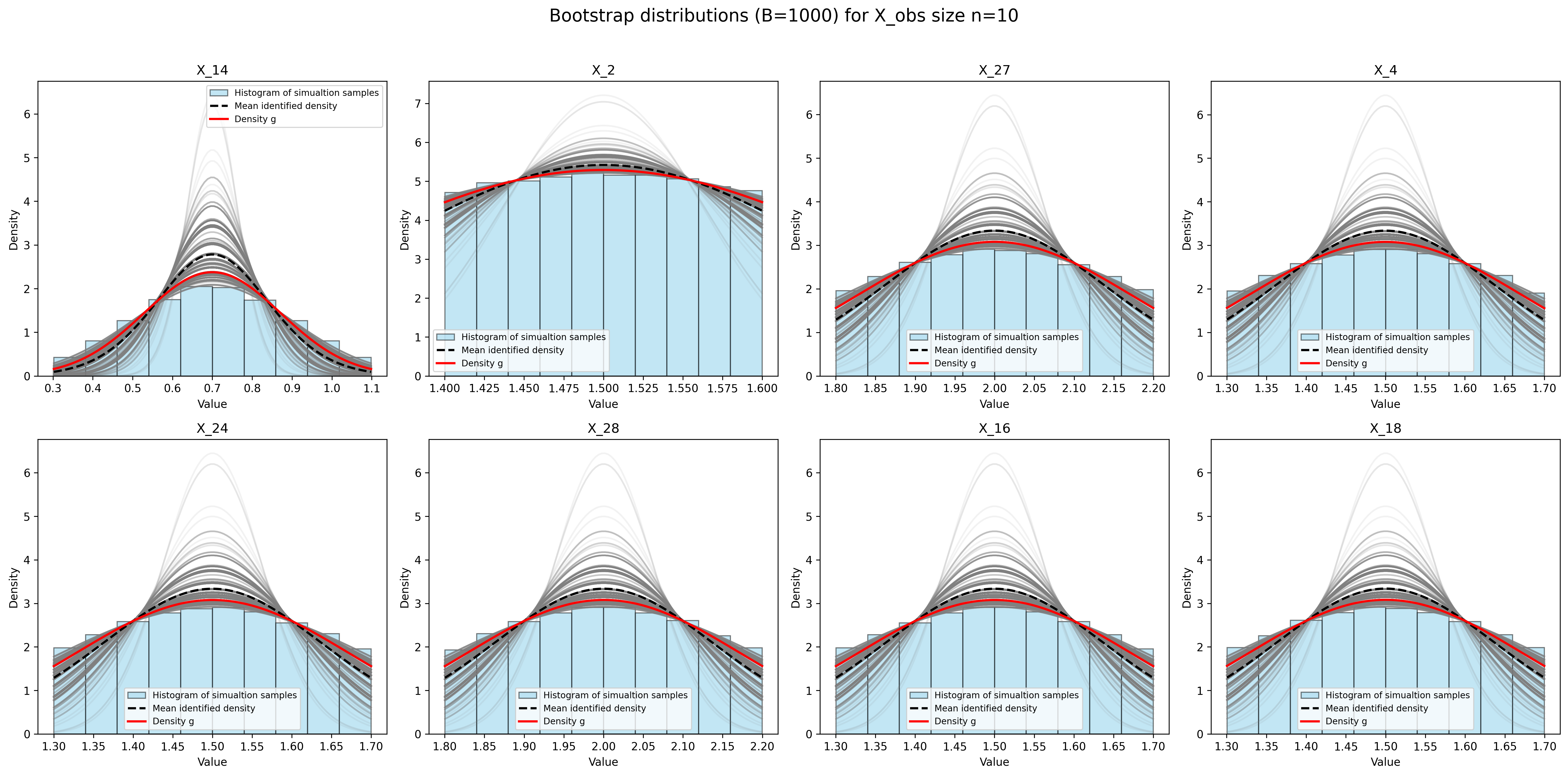}
    \caption{Effect of variability on the estimated marginal distributions for 8 selected gaps, showing gray curves for the identified distributions and the auxiliary distribution $g$ in red.}
    \label{fig:marginal}
\end{figure}

Figure~\ref{fig:marginal} illustrates the effect of epistemic uncertainty introduced by estimating input variances from a limited dataset of only 10 measurements ($N_{\text{test}}=10$) for only 8 variables. The figure displays three distinct elements: the red curve represents the auxiliary density $g$ used for Importance Sampling; the blue histogram shows the distribution of the 999 simulation samples drawn according to $g$; and the gray curves depict the family of distributions identified via bootstrap resampling of the experimental data.Although the dataset consists of only 10 measurements, a bootstrap of size 1000 was performed to approximate the distribution of the sample variances. We acknowledge that resampling from such a small dataset limits the fidelity of the bootstrap approximation and may lead to an overly optimistic representation of variability. Nonetheless, given the absence of additional measurements and the need to quantify epistemic uncertainty on the input distributions, this procedure provides a non-parametric approach for exploring the plausible range of variance estimates. It therefore offers an informative—albeit approximate—characterization of the uncertainty induced by limited data.

The output variables are the transmitted forces \(\mathbf{F} = (F_1, \ldots, F_{m})\), computed during the aircraft loading phases. The analysis focuses on a single critical interface, specifically the second interface, denoted as \(\phi = F_2\).

A total of 999 numerical simulations were performed following a fixed density \(
g_{X_i} = \mathcal{N}_{[a_i,\, b_i]}(\mu_i, {\sigma_i^*}^2),
\quad \text{for } i = 1, \dots, m.
\) The objective is to propagate epistemic uncertainties related to input law identification, output estimation, and the use of a Gaussian process, in order to estimate the B-basis value. In this case, the B-basis value corresponds to the upper bound of the 95\,\% CI of the 10\,\% quantile of the transmitted forces.

\begin{figure}[H]
    \centering
\includegraphics[width=1.25\textwidth]{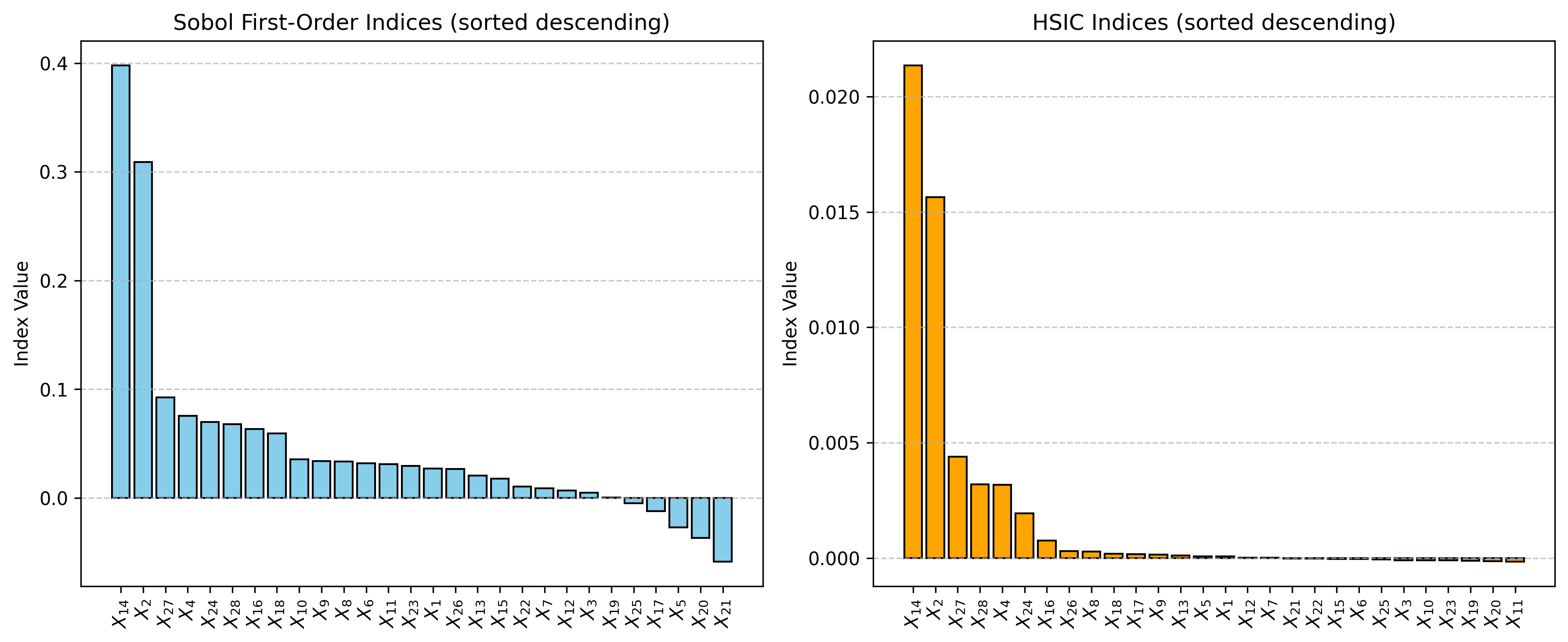}
    \caption{Sobol and HSIC sensitivity analysis}
    \label{fig:Sobol_Hsic}
\end{figure}

A Sobol \cite{Sobol1993} and HSIC \cite{Gretton2005} sensitivity analysis was performed (See Figure  \ref{fig:Sobol_Hsic}) to reduce the dimension of the input variables with respect to \(F_2\). This analysis identified the most influential independent variables: \\
\((X_{14}, X_2, X_{27}, X_4, X_{24}, X_{28}, X_{16}, X_{18})\), which will be the focus of the subsequent analysis.

\subsubsection{Results}
The results were obtained using a Monte Carlo simulation with a fixed budget $N_{\text{sim}}=649$. We compare different scenarios based on the quantity of physical tests ($N_{\text{test}}$) and the quality of the surrogate model ($N_{\text{DoE}}$).

\begin{figure}[H]
    \centering
    \includegraphics[width=1.15\textwidth]{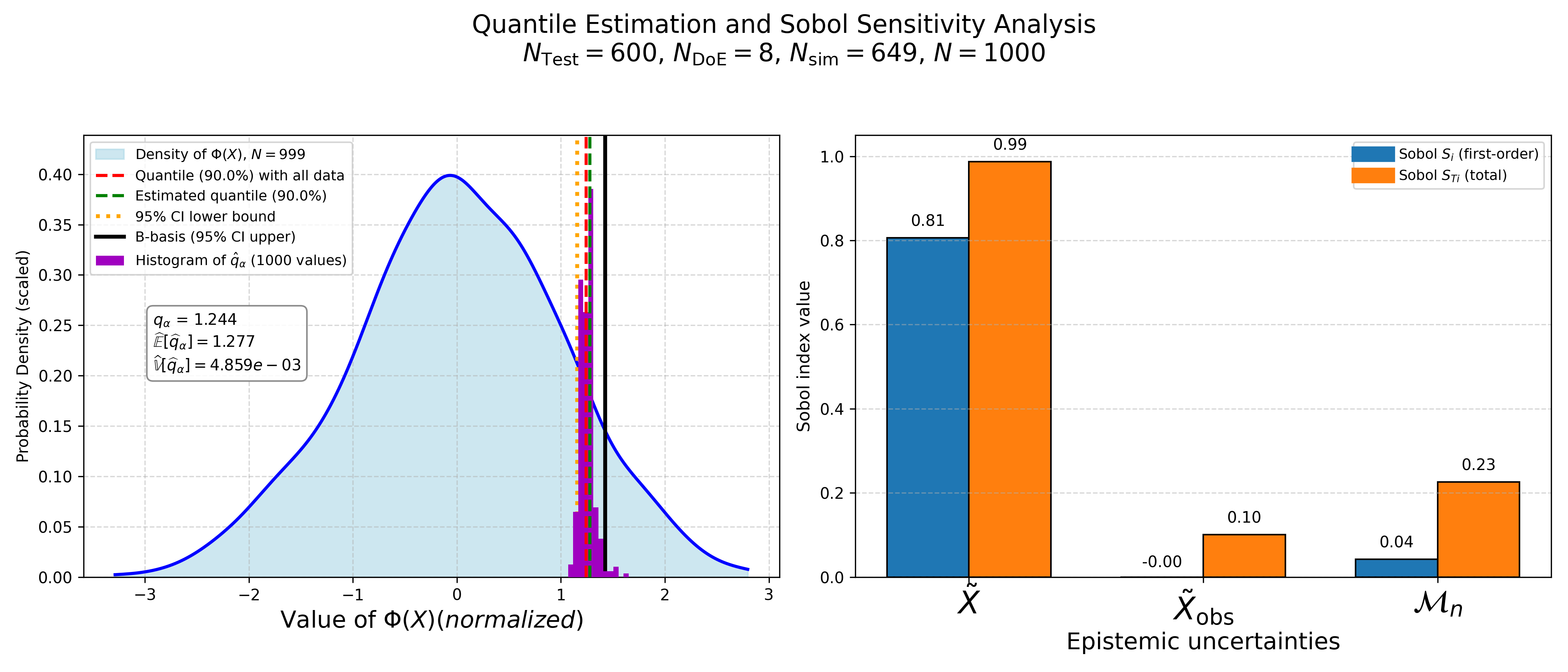}
    \caption{Result for Monte Carlo sampling uncertainties only, good input identification and poor surrogate model. 
    \textbf{Left:} histogram of the estimated quantile and the corresponding B-basis; 
    \textbf{Right:} first-order and total Sobol indices for all sources of epistemic uncertainty.}
    \label{result:fig4}
\end{figure}

\begin{figure}[H]
    \centering
    \includegraphics[width=1.15\textwidth]{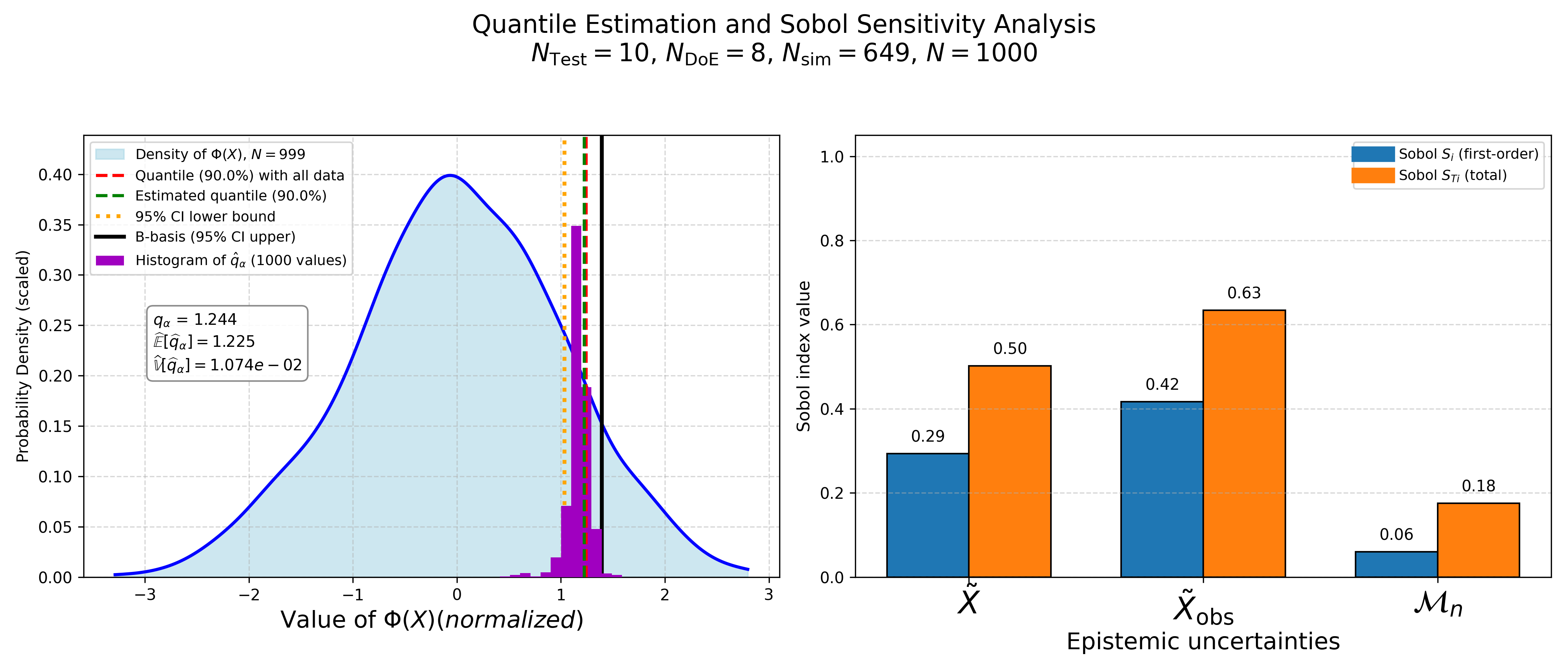}
    \caption{Result for Monte Carlo sampling uncertainties, bad input identification and poor surrogate model. 
    \textbf{Left:} histogram of the estimated quantile and the corresponding B-basis; 
    \textbf{Right:} first-order and total Sobol indices for all sources of epistemic uncertainty.}
    \label{result:fig5}
\end{figure}

\begin{figure}[H]
    \centering
    \includegraphics[width=1.15\textwidth]{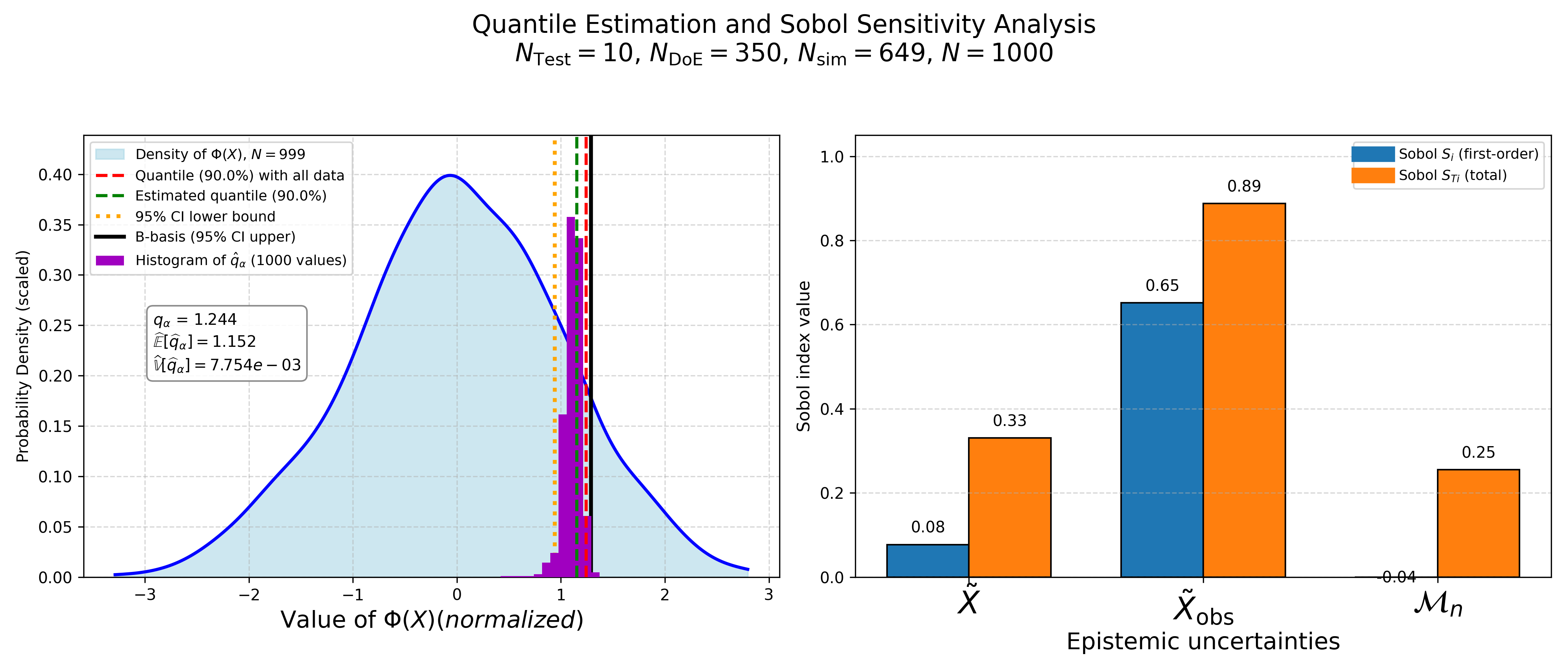}
    \caption{Result for Monte Carlo sampling uncertainties, bad input identification and good surrogate model.
    \textbf{Left:} histogram of the estimated quantile and the corresponding B-basis; 
    \textbf{Right:} first-order and total Sobol indices for all sources of epistemic uncertainty.}
    \label{result:fig6}
\end{figure}

\begin{figure}[H]
    \centering
    \includegraphics[width=1.15\textwidth]{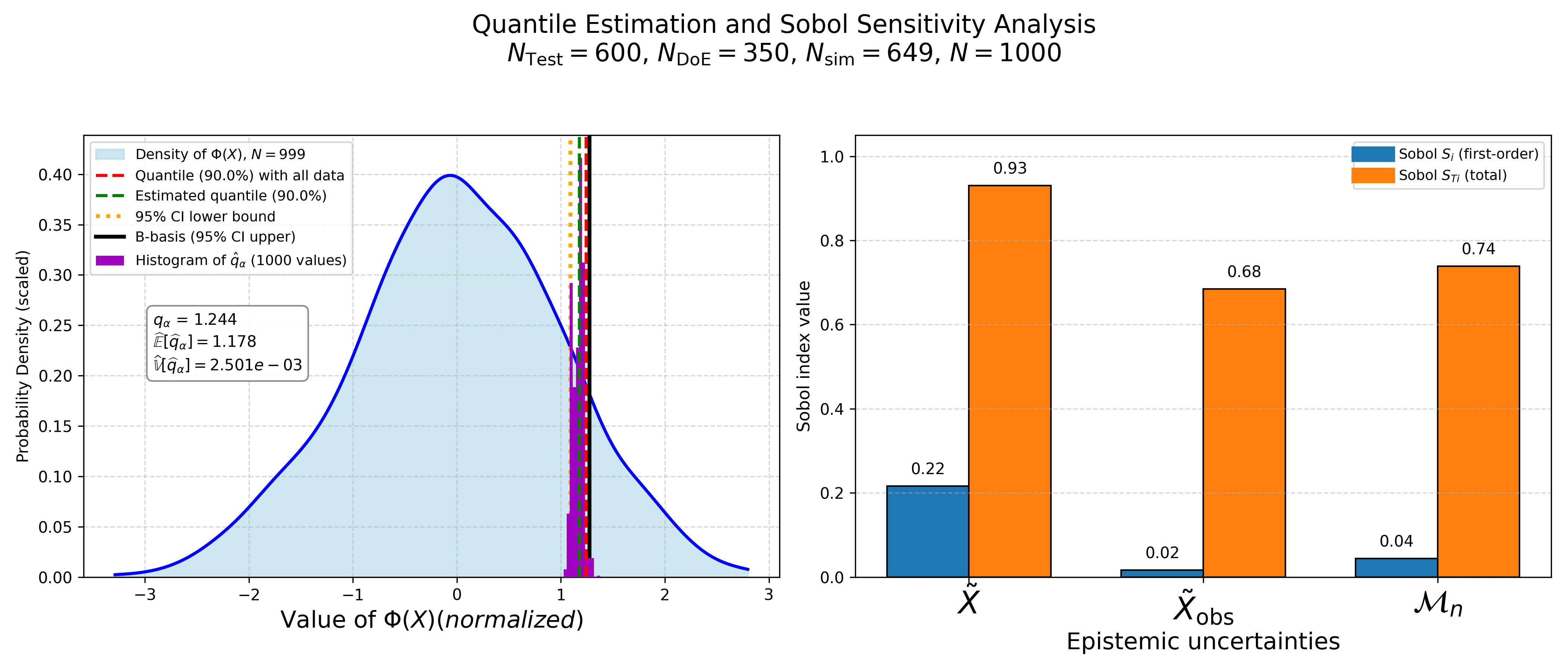}
    \caption{Result for Monte Carlo sampling uncertainties, good input identification and good surrogate model.
    \textbf{Left:} histogram of the estimated quantile and the corresponding B-basis; 
    \textbf{Right:} first-order and total Sobol indices for all sources of epistemic uncertainty.}
    \label{result:fig7}
\end{figure}

Figure \ref{result:fig5} illustrates a common industrial constraint: sparse physical data ($N_{\text{test}}=10$) combined with a poor surrogate. Despite these limitations, the method provides a conservative and safe B-basis estimate. This capability guarantees compliance early in the program, allowing the definition of numerical design allowables without waiting for extensive test campaigns. The method effectively manages the risks associated with low $N_{\text{test}}$.

Comparing Figure \ref{result:fig5} (poor surrogate, $N_{\text{DoE}}=8$) and Figure \ref{result:fig6} (refined surrogate, $N_{\text{DoE}}=350$) highlights the robustness of the approach. While a better model reduces variance and tightens the confidence interval, a poor surrogate does not introduce bias. The Sobol indices confirm that the impact of the surrogate model ($\mathcal{M}$) on the estimator's variance remains consistently negligible (near zero). This decoupling is strictly controlled by the Control Variates technique, which effectively cancels out the surrogate's error. This allows the safe deployment of fast surrogate models with no risk of underestimating structural loads.

Finally, the Sobol indices (right side of the figures) serve as a strategic guide for budget allocation:
\begin{itemize}
    \item In Figure \ref{result:fig5} ($N_{\text{test}}=10$), uncertainty is dominated by input identification ($\mathbf{\widetilde{X}}_{\text{obs}}$). Investing in numerical calculation is useless here; physical testing must be increased.
    \item In Figure \ref{result:fig4} ($N_{\text{test}}=600$), uncertainty is driven by Monte Carlo sampling ($\mathbf{\widetilde{X}}$). In this case, increasing the number of simulations ($N_{\text{sim}}$) is the priority.
    \item Comparatively, Figure \ref{result:fig7} ($N_{\text{test}}=600, N_{\text{DoE}}=350$) represents the optimal scenario. By combining rich physical data with a high-fidelity surrogate, the total variance is minimized, leaving only the irreducible aleatory uncertainty.
\end{itemize}

This framework optimizes the investment by directing resources ($N_{\text{test}}$, $N_{\text{sim}}$ or $N_{\text{DoE}}$) only where they effectively reduce design margins. Applied to a future \text{Airbus commercial aircraft}, the method delivers a robust numerical B-basis even under high epistemic uncertainty, avoiding deterministic safety factors and streamlining the certification workflow.

\section{Conclusion}
\label{conclusion}
This study has addressed the challenge of estimating conservative design quantiles (A- and B-basis) under mixed \textit{aleatory} and \textit{epistemic} uncertainties using numerical models. By proposing a unified numerical stochastic framework, we have demonstrated that combining Importance Sampling with Control Variates allows for rigorous reliability assessment even when dealing with limited data and computationally expensive models.

The primary contribution of this work lies in the successful integration of surrogate models solely as variance reduction tools. Our numerical experiments confirm that this strategy preserves the unbiasedness and consistency of the estimator---a mandatory requirement for certification-while significantly reducing the computational budget compared to standard Monte Carlo methods. Consequently, this approach provides a statistically justified framework to estimate A-basis and B-basis using numerical model. Furthermore, the automatic derivation of Sobol indices offers valuable insights into the hierarchy of uncertainty sources. Moreover, the proposed propagation framework is agnostic to the specific method used for input distribution identification. While this study employed a non-parametric approach coupled with bootstrap resampling, the methodology naturally accommodates alternative strategies, such as parametric Bayesian inference, without requiring structural modifications.

Building on this versatility, a valuable perspective for future work would be to benchmark different identification methods and input modeling assumptions (e.g., independence versus correlation) to assess their respective impact on the conservatism of the final design allowables.
Despite these advancements, Sobol indices remain global tools and do not provide actionable guidance for surrogate model enrichment. This underscores the necessity for developing new sensitivity measures for adaptive enrichment strategies. Another promising direction is the adaptive design of auxiliary densities in importance sampling, in order to better balance accuracy and efficiency in reliability contexts. Finally, extending this variance reduction framework to Multilevel Monte Carlo strategies \cite{Mycek2019} presents a promising pathway to handle high-fidelity simulations in complex industrial assemblies.

Overall, this work sets the ground for a new class of uncertainty quantification methods that combine statistical rigor with practical efficiency, offering a credible path toward more rational and less conservative design practices in engineering.

\section*{Declaration of competing interest}
The authors declare that they have no known competing financial interests or personal relationships that could
have appeared to influence the work reported in this paper.
\section*{Declaration of Generative AI and AI-Assisted Technologies in the Manuscript Preparation Process}

During the preparation of this work, the authors used Gemini to refine the English language phrasing, correct errors, and assist with text formatting. The authors reviewed and edited the output as needed and take full responsibility for the content of the published article.

\appendix
\section{Definition and limit of the corrected CV-IS Weights Sum}
\label{Weigh_IS}
To derive the explicit form of the weights $W_j$, we first recall the definition of the optimal control coefficient $C(y)$, which minimizes the variance of the estimator. It is defined as the ratio of the covariance to the variance of the control variate.

In the empirical estimation, we leverage the following covariance identity for two random variables $U$ and $V$:
\begin{equation}
    \text{Cov}(U, V) = \mathbb{E}\left[(U - \mathbb{E}[U])(V - \mathbb{E}[V])\right] = \mathbb{E}\left[U(V - \mathbb{E}[V])\right].
    \label{eq:cov_identity}
\end{equation}
By applying Eq.~\eqref{eq:cov_identity} to our problem, we identify the variables as follows:
\begin{itemize}
    \item $U = \mathbf{1}_{\phi(\mathbf{X}) \leq y} \lambda^{\text{IS}}(\mathbf{X})$ (the target variable),
    \item $V = \mathbf{1}_{\mathcal{M}(\mathbf{X}) \leq z_\alpha} \lambda^{\text{IS}}(\mathbf{X})$ (the control variable),
    \item $\mathbb{E}[V] \approx \widehat{h}$ (the empirical mean).
\end{itemize}

Consequently, the empirical numerator for $\widehat{C}(y)$ simplifies to $\sum_{j} U_j (V_j - \widehat{h})$. 

Substituting this into the CV-IS cdf estimator expansion yields:
\begin{align*}
   \widehat{F_{Y}^{\text{CV-IS}}}(y) & = \widehat{F^{\text{IS}}_Y} - \widehat{C}(y)(\widehat{h} - \mathbb{E}_f[\mathbf{1}_{\mathcal{M}(\mathbf{X}) \leq z_\alpha}])\\
  & = \widehat{F^{\text{IS}}_Y} - \widehat{C}(y)(\widehat{h} - \alpha)\\
  & = \widehat{F^{\text{IS}}_Y} - \frac{\sum_{j=1}^{N_{\text{sim}}}\mathbf{1}_{\phi(\mathbf{X}_j) \leq y}\lambda^{\text{IS}} (\mathbf{X}_j) (\mathbf{1}_{\mathcal{M}(\mathbf{X}_j) \leq z_\alpha}\lambda^{\text{IS}} (\mathbf{X}_j) - \widehat{h})}{\sum_{i=1}^{N_{\text{sim}}} (\mathbf{1}_{\mathcal{M}(\mathbf{X}_i) \leq z_\alpha}\lambda^{\text{IS}} (\mathbf{X}_i) - \widehat{h})^2} (\widehat{h} -  \alpha)\\
  & = \sum_{j=1}^{N_{\text{sim}}} \frac{1}{N_{\text{sim}}}\mathbf{1}_{\phi(\mathbf{X}_j) \leq y}\lambda^{\text{IS}} (\mathbf{X}_j) -\sum_{j=1}^{N_{\text{sim}}}\mathbf{1}_{\phi(\mathbf{X}_j) \leq y}\lambda^{\text{IS}} (\mathbf{X}_j) \frac{ (\mathbf{1}_{\mathcal{M}(\mathbf{X}_j) \leq z_\alpha}\lambda^{\text{IS}} (\mathbf{X}_j) - \widehat{h})}{\sum_{i=1}^{N_{\text{sim}}} (\mathbf{1}_{\mathcal{M}(\mathbf{X}_i) \leq z_\alpha}\lambda^{\text{IS}} (\mathbf{X}_i) - \widehat{h})^2} (\widehat{h} -  \alpha) \\
  & =\sum_{j=1}^{N_{\text{sim}}} \left[ \frac{1}{N_{\text{sim}}} - \frac{ (\widehat{h} - \alpha) (\mathbf{1}_{\mathcal{M}(\mathbf{X}_j) \leq z_\alpha}\lambda^{\text{IS}} (\mathbf{X}_j) - \widehat{h})}{\sum_{i=1}^{N_{\text{sim}}} (\mathbf{1}_{\mathcal{M}(\mathbf{X}_i) \leq z_\alpha}\lambda^{\text{IS}} (\mathbf{X}_i) - \widehat{h})^2} \right] \lambda^{\text{IS}} (\mathbf{X}_j) \mathbf{1}_{\phi(\mathbf{X}_j) \leq y} \\
    &=\sum_{j=1}^{N_{\text{sim}}} W_j \mathbf{1}_{\phi(\mathbf{X}_j) \leq y}          
\end{align*}
Finally by identification:
\begin{equation}
  W_j =\left[ \frac{1}{N_{\text{sim}}} + \frac{ (\widehat{h} - \alpha) (\widehat{h}-\mathbf{1}_{\mathcal{M}(\mathbf{X}_j) \leq z_\alpha}\lambda^{\text{IS}}(\mathbf{X}_j) )}{\sum_{i=1}^{N_{\text{sim}}} (\mathbf{1}_{\mathcal{M}(\mathbf{X}_i) \leq z_\alpha}\lambda^{\text{IS}}(\mathbf{X}_i) - \widehat{h})^2} \right] \lambda^{\text{IS}} (\mathbf{X}_j) 
\end{equation}

We consider the limit, as \(n \ (N_{\text{sim}} = n) \to +\infty\), of the following expression:
\begin{equation}
\sum_{i=1}^{n} W_i = \frac{1}{n} \sum_{i=1}^{n} \lambda^{\text{IS}}(\mathbf{X}_i) 
- \frac{(\widehat{h} - \alpha)}{\sum_{i=1}^{n} (\mathbf{1}_{\mathcal{M}(\mathbf{X}_i) \leq z_\alpha} \lambda^{\text{IS}}(\mathbf{X}_i) - \widehat{h})^2}
\left( \sum_{i=1}^{n} \left( \mathbf{1}_{\mathcal{M}(\mathbf{X}_i) \leq z_\alpha} \lambda^{\text{IS}}(\mathbf{X}_i)^2 - \widehat{h} \lambda^{\text{IS}}(\mathbf{X}_i) \right) \right),
\end{equation}
where:
\begin{itemize}
  \item \( \lambda^{\text{IS}}(\mathbf{X}) = \frac{f_X(\mathbf{X})}{g(\mathbf{X})} \) is the importance weight,
  \item \( \widehat{h} = \frac{1}{n} \sum_{i=1}^{n} \mathbf{1}_{\mathcal{M}(\mathbf{X}_i) \leq z_\alpha} \lambda^{\text{IS}}(\mathbf{X}_i) \),
  \item \( z_\alpha \) is chosen such that: \( \mathbb{E}_f \left[ \mathbf{1}_{\mathcal{M}(\mathbf{X}) \leq z_\alpha}  \right] =
  \mathbb{E}_g \left[ \mathbf{1}_{\mathcal{M}(\mathbf{X}) \leq z_\alpha} \lambda^{\text{IS}}(\mathbf{X}) \right]= \alpha.
  \)
\end{itemize}

\paragraph{Step 1: Behavior of the first term.}

By the strong law of large numbers applied to the i.i.d. and integrable variables \( \lambda^{\text{IS}}(\mathbf{X}_i) \), we have:
\begin{equation}
\frac{1}{n} \sum_{i=1}^{n} \lambda^{\text{IS}}(\mathbf{X}_i) \xrightarrow[n \to +\infty]{a.s.} \mathbb{E}_g[\lambda^{\text{IS}}(\mathbf{X})] = 1.
\end{equation}

\paragraph{Step 2: Behavior of the second term.}

Now consider the second term:
\begin{align*}
T_n &= \frac{(\widehat{h} - \alpha)}{\sum_{i=1}^{n} (\mathbf{1}_{\mathcal{M}(\mathbf{X}_i) \leq z_\alpha} \lambda^{\text{IS}}(\mathbf{X}_i) - \widehat{h})^2}
\left( \sum_{i=1}^{n} \left( \mathbf{1}_{\mathcal{M}(\mathbf{X}_i) \leq z_\alpha} \lambda^{\text{IS}}(\mathbf{X}_i)^2 - \widehat{h} \lambda^{\text{IS}}(\mathbf{X}_i) \right) \right) \\
&= \frac{(\widehat{h} - \alpha)}{D_n} N_n,
\end{align*}
where
\(
D_n = \sum_{i=1}^{n} (\mathbf{1}_{\mathcal{M}(\mathbf{X}_i) \leq z_\alpha} \lambda^{\text{IS}}(\mathbf{X}_i) - \widehat{h})^2,
\)
and
\(
N_n = \sum_{i=1}^{n} \left( \mathbf{1}_{\mathcal{M}(\mathbf{X}_i) \leq z_\alpha} \lambda^{\text{IS}}(\mathbf{X}_i)^2 - \widehat{h} \lambda^{\text{IS}}(\mathbf{X}_i) \right).
\)

\paragraph{Step 3: Limit of the denominator}
Dividing \(D_n\) by \(n\), the strong law of large numbers gives:
\[
\frac{D_n}{n} \xrightarrow[n \to +\infty]{a.s.} \mathbb{V}_g\big(\mathbf{1}_{\mathcal{M}(\mathbf{X}) \leq z_\alpha} \lambda^{\text{IS}}(\mathbf{X})\big) > 0.
\]

\paragraph{Step 4: Limit of the numerator.}

We rewrite the numerator as:
\[
N_n = \sum_{i=1}^{n} \mathbf{1}_{\mathcal{M}(\mathbf{X}_i) \leq z_\alpha} \lambda^{\text{IS}}(\mathbf{X}_i)^2 - \widehat{h} \sum_{i=1}^{n} \lambda^{\text{IS}}(\mathbf{X}_i).
\]
By the strong law of large numbers:
\[
\frac{1}{n} \sum_{i=1}^{n} \mathbf{1}_{\mathcal{M}(\mathbf{X}_i) \leq z_\alpha} \lambda^{\text{IS}}(\mathbf{X}_i)^2 \xrightarrow[n \to +\infty]{a.s.} \mathbb{E}_g \left[ \mathbf{1}_{\mathcal{M}(\mathbf{X}) \leq z_\alpha} \lambda^{\text{IS}}(\mathbf{X})^2 \right],
\]
and
\[
\frac{1}{n} \sum_{i=1}^{n} \lambda^{\text{IS}}(\mathbf{X}_i) \xrightarrow[n \to +\infty]{a.s.} 1,
\]
while \( \widehat{h} \to \alpha \). Hence,
\[
\frac{N_n}{n} \xrightarrow[n \to +\infty]{a.s.} \mathbb{E}_g \left[ \mathbf{1}_{\mathcal{M}(\mathbf{X}) \leq z_\alpha} \lambda^{\text{IS}}(\mathbf{X})^2 \right] - \alpha.
\]

\paragraph{Step 5: Limit of \(T_n\)}
Recall:
\[
T_n = \frac{(\widehat{h} - \alpha)}{D_n} N_n.
\]
Since
\[
\widehat{h} - \alpha \xrightarrow[n \to +\infty]{a.s.} 0,
\quad
\frac{D_n}{n} \xrightarrow[n \to +\infty]{a.s.} \mathbb{V}_g\big(\mathbf{1}_{\mathcal{M}(\mathbf{X}) \leq z_\alpha} \lambda^{\text{IS}}(\mathbf{X})\big) > 0, \] and\[
\frac{N_n}{n} \xrightarrow[n \to +\infty]{a.s.} \mathbb{E}_g \left[ \mathbf{1}_{\mathcal{M}(\mathbf{X}) \leq z_\alpha} \lambda^{\text{IS}}(\mathbf{X})^2 \right] - \alpha,
\]
we obtain
\[
T_n = \frac{o(1)}{O(n)} \times O(n) = o(1),
\]
which implies
\[
T_n \xrightarrow[n \to +\infty]{a.s.} 0.
\]
Finally, we conclude that
\[
\sum_{i=1}^{n} W_i \xrightarrow[n \to +\infty]{a.s.} 1.
\]

\bibliographystyle{elsarticle-num} 
\bibliography{Mabib}





\end{document}